\documentclass[prl, twocolumn,superscriptaddress,amsmath,amssymb,showpacs,floatfix,preprintnumbers]{revtex4-2}

\usepackage{amsmath}
\usepackage{graphicx}
\usepackage{dcolumn}
\usepackage{xcolor}
\usepackage{physics}
\usepackage{xr}
\usepackage{siunitx}
\usepackage{tabularx}
\usepackage{makecell}
\usepackage{multirow}
\DeclareGraphicsExtensions{.png .jpg .pdf}

\usepackage[normalem]{ulem}

\makeatletter
\newcommand*{\addFileDependency}[1]{
  \typeout{(#1)}
  \@addtofilelist{#1}
  \IfFileExists{#1}{}{\typeout{No file #1.}}
}
\makeatother
 
\newcommand*{\myexternaldocument}[1]{%
    \externaldocument{#1}%
    \addFileDependency{#1.tex}%
    \addFileDependency{#1.aux}%
}

\myexternaldocument{si}

\begin{document}

\title{Ferroelectric Semimetals with $\alpha$-Bi/SnSe van der Waals heterostructures and its Topological Currents}

\author{D. J. P. de Sousa}
\affiliation{Department of Electrical and Computer Engineering, University of Minnesota, Minneapolis, Minnesota 55455, USA}
\author{Seungjun Lee}
\affiliation{Department of Electrical and Computer Engineering, University of Minnesota, Minneapolis, Minnesota 55455, USA}
\author{Qiangsheng Lu}
\affiliation{Materials Science and Technology Division, Oak Ridge National Laboratory, Oak Ridge, Tennessee 37831, USA}
\author{Rob G. Moore}
\affiliation{Materials Science and Technology Division, Oak Ridge National Laboratory, Oak Ridge, Tennessee 37831, USA}
\author{Matthew Brahlek}
\affiliation{Materials Science and Technology Division, Oak Ridge National Laboratory, Oak Ridge, Tennessee 37831, USA}
\author{J-.P. Wang}
\affiliation{Department of Electrical and Computer Engineering, University of Minnesota, Minneapolis, Minnesota 55455, USA}
\affiliation{Department of Physics, University of Minnesota, Minneapolis, Minnesota 55455, USA}
\author{Guang Bian}\email{biang@missouri.edu}
\affiliation{Department of Physics and Astronomy, University of Missouri, Columbia, Missouri 65211, USA}
\author{Tony Low}\email{tlow@umn.edu}
\affiliation{Department of Electrical and Computer Engineering, University of Minnesota, Minneapolis, Minnesota 55455, USA}
\affiliation{Department of Physics, University of Minnesota, Minneapolis, Minnesota 55455, USA}

\begin{abstract}
We show that proximity effects can be utilized to engineer van der Waals heterostructures (vdWHs) displaying spin-ferroelectricity locking, where ferroelectricity and spin states are confined to different layers, but are correlated by means of proximity effects. Our findings are supported by first principles calculations in $\alpha$-Bi/SnSe bilayers. We show that such systems support ferroelectrically switchable non-linear anomalous Hall effect originating from large Berry curvature dipoles as well as direct and inverse spin Hall effects with giant bulk spin-charge interconversion efficiencies. The giant efficiencies are consequences of the proximity-induced semimetallic nature of low energy electron states, which are shown to behave as two-dimensional pseudo-Weyl fermions by means of symmetry analysis, first principles calculations as well as direct angle-resolved photoemission spectroscopy 
 measurements.

\end{abstract}

\maketitle

\emph{Introduction.}\textemdash 
Proximity effects in van der Waals heterostructures (vdWHs) have enabled unprecedented control over the electronic structure and spin degree-of-freedom of electrons in two-dimensional (2D) crystal lattices~\cite{Huang2020, PhysRevLett.119.146401, PhysRevB.104.195156, PhysRevLett.125.196402, Bora2021, proximity1, proximity2, proximity3, giantspinlifetime1, giantspinlifetime2, topproximity, Avsar2014}. Unique 2D electron states with distinctive spin textures can be engineered by combining atomically thin layers displaying varying degrees of spin-orbit and/or exchange interactions, an approach that has been successfully shown to induce alternative current responses, including anomalous Hall effects and spin-charge interconversion (SCI) in systems that otherwise would not support such features~\cite{PhysRevLett.114.016603, PhysRevB.103.125304, CSC1, CSC2, CSC3, CSC4, PhysRevB.106.165420, Garcia2017}. A relatively unexplored direction is the possibility to proximitize atomically thin ferroelectrics with 2D materials displaying strongly spin-orbit coupled electron states to induce correlations between ferroelectric polarization and spins. As a result of this ferroelectric control of the spin-texture~\cite{Rinaldi2018}, or simply spin-ferroelectricity locking, unique 2D states with electrically tunable unusual current responses are expected to exist, which holds the promise to enable next-generation electronic devices~\cite{Manipatruni2018, https://doi.org/10.48550/arxiv.2302.12162, Mankalale2017}.

While switchable spin Hall currents through spin-ferroelectricity locking in three-dimensional (3D) ferroelectric Rashba semiconductors have been demonstrated experimentally~\cite{2Zhang2020, Wang2020, Varotto2021}, the effect remains elusive in semimetallic systems, which are expected to host giant SCI coefficients due to concomitant Berry curvature hotspots and suppressed density of states at semimetallic band crossings~\cite{Zhang2022, PhysRevB.97.041101, Liu2018}. Ferroelectric semimetals displaying giant Berry curvature dipoles (BCDs) are also expected to support highly efficient generation of non-linear anomalous Hall currents, which also remains a relatively unexplored phenomenon due to the lack of appropriate material candidates~\cite{Zhang2022}. In this respect, the possibility of achieving semimetallic states with spin-ferroelectricity locking  represents an crucial step towards the realization and study of such an unique and unusual current responses. 

In this work, we show that 2D semimetallic states displaying spin-ferroelectricity locking can be engineered by proximitizing $\alpha$-Bi, a black phosphorus-like monolayer of Bi atoms~\cite{Kowalczyk2020}, onto SnSe(S), an atomically thin ferroelectric with in-plane polarization~\cite{Higashitarumizu2020, Chang2020, Du2022, PhysRevB.101.184101, Wang2017, Bao2019}. The semimetallic nature of such vdWHs have been directly accessed through angle-resolved photoemission spectroscopy (ARPES) measurements and is well described by our first principles calculations and symmetry analysis. We also predict that such vdWHs host giant BCDs and SCI coefficients, paving the way for the efficient ferroelectric control of spin Hall and non-linear Hall effects.

\begin{figure*}[t]
\centerline{\includegraphics[width = 0.8\linewidth]{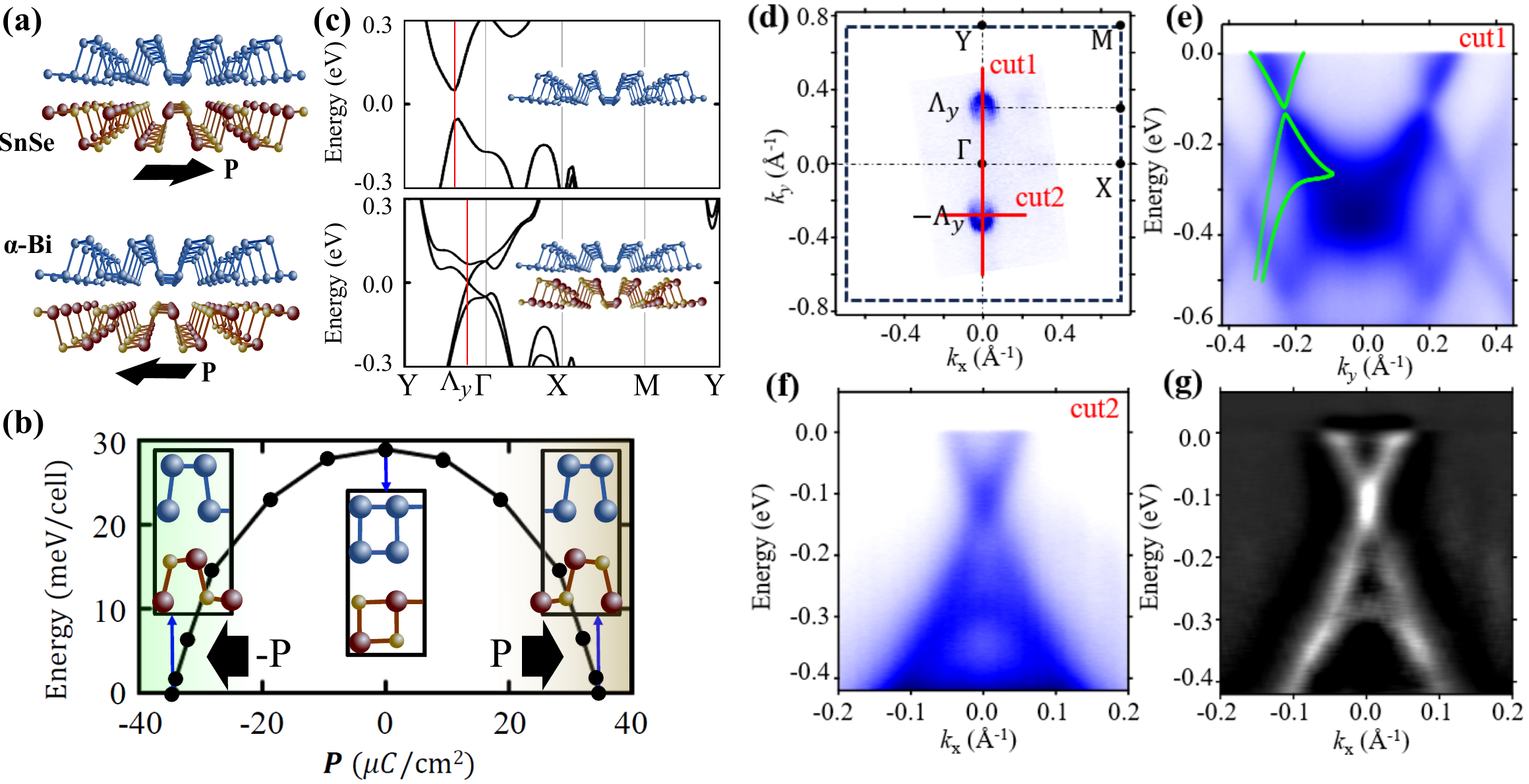}}
\caption{$\alpha$-Bi/SnSe van der Waals heterostructure. Two distinct in-plane polarization states are depicted in (a). The energy barrier between the two polarization states and corresponding spontaneous polarizations of $\alpha$-Bi/SnSe are shown in (b). Proximity with the in-plane polarization of the SnSe substrate breaks the degeneracy of low energy $p_z$ states of $\alpha$-Bi, inducing a gap closure. The band structures of free-standing $\alpha$-Bi and $\alpha$-Bi/SnSe vdwHs as obtained from first principles are show in (c), where the proximity induced gap closure occuring at the $\Lambda_y$ point becomes explicit. (d) Fermi surface of $\alpha$-Bi/SnSe mapped by ARPES. (e) ARPES spectrum taken along the line of ``cut1” marked in (d). The red curves are calculated band dispersion along $\Gamma$-Y direction. The green curves are bands along $\Gamma$-X from the 90{$^\circ$}-rotated domains in the MBE sample. (f) ARPES spectrum taken along ``cut2”. (d) Second derivative of the spectrum in (g) for a better visualization of the band dispersion. 
}
\label{Fig1}
\end{figure*}

\emph{Crystal and electronic structures of $\alpha$-Bi/SnSe} \textemdash
 We begin by describing crystal structure of $\alpha$-Bi/SnSe vdWHs as schematically shown in Fig.~\ref{Fig1}(a). $\alpha$-Bi has an inversion symmetry and is classified into $Cmce$ space group which belongs to the same space group of black phosphorus (BP). Although small buckling in the Bi-Bi bond lowers the total energy of a freestanding $\alpha$-Bi monolayer, $Cmce$ $\alpha$-Bi can be stabilized by carrier doping~\cite{Lu2015} and was realized experimentally~\cite{Kowalczyk2020}. 
More recently, Lu et al~\cite{lu2023observation} found that  $Cmce$ $\alpha$-Bi is spontaneously stabilized on group-IV monochalcogenides substrates and exhibit unique 2D pseudo Weyl states. Here, we note that there are two energetically equivalent $\alpha$-Bi/group-IV monochalcogenides vdWHs, which are mirror counterparts and thus show opposite in-plane dipole moments intrinsically originating from group-IV monochalcogenides, as depicted in Fig.~\ref{Fig1}(a).

To understand the role of the ferroelectric group-IV monochalcogenides substrate, we performed first-principles density functional theory calculations for $\alpha$-Bi/SnSe vdWHs as an exemplary heterostructure~(See Supplemental Material for details). We found that the band gap of $\alpha$-Bi is almost closed when it is on the SnSe substrate, which well agrees with the previous study~\cite{lu2023observation}. Then, we performed nudged elastic band (NEB) calculation~\cite{henkelman2000climbing} to evaluate the energy barrier ($E_{\rm{B}}$) between two equivalent vdWHs and their spontaneous polarizations $P_{\rm{s}}$, as shown in Fig.~\ref{Fig1}(b). $E_{\rm{B}}$ and $P_{\rm{s}}$ were evaluated to be 28.9~meV/cell and 34.48~${\mu}$C/cm$^2$, which are slightly higher value than those of monolayer SnSe ($E_{\rm{B}}$=7~meV/cell, $P_{\rm{s}}$=18.1~${\mu}$C/cm$^2$)~\cite{Wang_2017}~(See Supplemental Material for more details). This clearly validates that $\alpha$-Bi/SnSe vdWHs also exhibits stable $P_{\rm{s}}$. 

Figure~\ref{Fig1}(c) displays the band structures of monolayer $\alpha$-Bi and $\alpha$-Bi/SnSe as obtained from our first principle calculations~\cite{Snote}. The presence of the SnSe substrate breaks the degeneracy of the low energy states of \textit{Cmce} $\alpha$-Bi around the $\Gamma$ point and induces an accidental gap closure with two-fold degenerate band crossings at momentum $\pm \Lambda_y$ along the $\Gamma$ to $\pm Y$ path. This result is consistent with the ARPES measurements shown in Fig.~\ref{Fig1}(d)~\cite{lu2023observation}. Bismuth was deposited on the cleaved surface of single crystal SnSe. The SnSe crystals are n-type doped with Br. The base pressure was lower than $2 \times 10^{-10}$ mbar. The temperature of the SnSe substrate was kept at 50 C$^\circ$ during the growth. The ARPES measurements were performed in a lab-based system coupled to the molecular beam epitaxy system, using a Scienta DA30L hemispherical analyzer with a base pressure of $< 5 \times 10^{-11}$ mbar and a base temperature of $T \approx 8$ K. The light source for ARPES is an Oxide $11$ eV laser system. The energy resolution of ARPES measurements is $\approx 3$ meV. The ARPES Fermi surface and band spectra from $\alpha$-Bi/SnSe are shown in Fig.~\ref{Fig1}(d)-(g). The spectra along ``cut1” and ``cut2”, shown in Figs.~\ref{Fig1}(e) and (f), are in good agreement with first-principles results, indicating a gapless spin-valley polarized band dispersion. The second derivative spectrum highlights in finer details the semimetallic band crossing in Figs.~\ref{Fig1}(g). We note that the crossing bands describe low energy electron states confined to the $\alpha$-Bi layer. Thus, the conduction properties of the $\alpha$-Bi/SnSe vdWHs is purely dictated by these strongly spin-orbit coupled 2D massless pseudo Weyl fermions, giving rise to salient features that will be discussed in upcoming sections.

These observations make $\alpha$-Bi/SnSe vdWHs even more exotic because it implies that the strongly spin-orbit coupled pseudo Weyl states in the $\alpha$-Bi layer is coupled to the ferroelectricity of the SnSe substrate, which can be further manipulated by means of external electric fields. Hence, $\alpha$-Bi/SnSe vdWHs must display semimetallic spin-ferroelectricity locking, where ferroelectricity and spin transport are confined to different layers, but are correlated by means of proximity effects. In the following, we address the question of how proximity effects give rise to pseudo Weyl states in the $\alpha$-Bi layer and correlates it to the ferroelectricity of the SnSe substrate~\cite{Snote}.

\begin{figure}[t]
\centerline{\includegraphics[width = \linewidth]{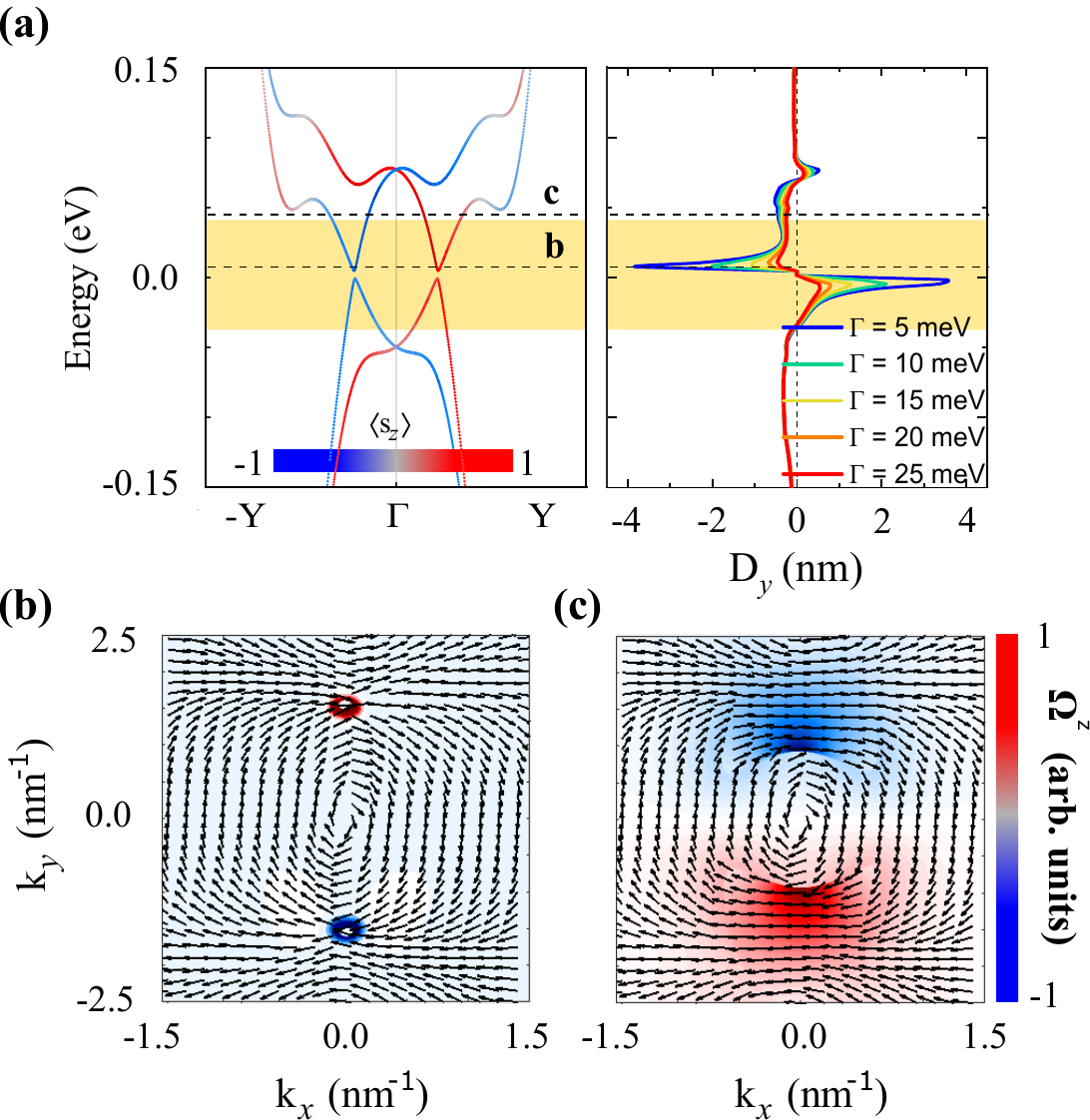}}
\caption{Berry curvature dipole (BCD) in $\alpha$-Bi/SnSe vdWHs. (a) Spin-resolved low energy bands of $\alpha$-Bi/SnSe obtained from first principles (left). The shaded region highlights the energy window within which the pseudo-Weyl description of electrons becomes relevant. The energy dependent BCD at several spectral broadenings $\Gamma$ is also shown (right). The BCD reaches it maximum within the pseudo-Weyl states energy window. (b) and (c) display the momentum-resolved Berry curvature field at the two energies highlighted by the horizontal dashed lines in panel (a).}
\label{Fig2}
\end{figure}

\emph{$\alpha$-Bi/SnSe: insulator-to-semimetal transition.} In $\alpha$-Bi/SnSe-type heterostructures, the long-wavelength physics is dominated by the $p_z$ states in the $\alpha$-Bi layer~\cite{Li2021}. The fully proximitized Hamiltonian for $p_z$ electrons in $\alpha$-Bi is $H(\textbf{k}) = H_{\textrm{$\alpha$-Bi}}(\textbf{k}) + H_{\textrm{prox}}$, where $H_{\textrm{$\alpha$-Bi}}(\textbf{k})$ ($H_{\textrm{prox}}$) describes the freestanding $\alpha$-Bi states (proximity effects). The freestanding $\alpha$-Bi $p_z$ states form the representation: 
$D(C_{2y}) = i\sigma_x s_x, D(C_{2z}) = -i\sigma_x s_z,  D(\mathcal{P}) = -i\sigma_x s_0, D(\mathcal{T}) = -i\sigma_0 s_y \mathcal{K}$, where $C_{2z(y)}$ is the twofold rotation around the the $z(y)$ axis, $\mathcal{P}$ is the inversion operator and $\mathcal{T}$ the time-reversal symmetry operator. Here, the $\sigma_{0, x, y, z} (s_{0, x, y, z})$ matrices operate in orbital (spin) space. Note that there is a mirror operator $\mathcal{M}_y$ whose representation is $D(\mathcal{M}_y) = D(C_{2z}C_{2y}) = -i\sigma_0 s_y $. The Hamiltonian constrained by these symmetries is black phosphorus-like~\cite{PhysRevB.96.155427, PhysRevB.94.235415, PhysRevB.92.075437, PhysRevLett.119.226801} with an additional Kane-Mele~\cite{PhysRevLett.95.226801} spin-orbit coupling term 
\begin{eqnarray}
&H_{\textrm{$\alpha$-Bi}}(\textbf{k}) = h_0(\textbf{k}) \sigma_0 s_0 + h_x(\textbf{k}) \sigma_x s_0+ h_y(\textbf{k}) \sigma_y s_0 + \nonumber \\ &h_{\textrm{SOC}}(\textbf{k}) \sigma_z s_z,
\label{eq2}
\end{eqnarray}
where $h_0(\textbf{k})=u_0 + \eta_x k_x^2 + \eta_y k_y^2$, $h_x(\textbf{k}) = \delta_0 + \gamma_x k_x^2 + \gamma_y k_y^2$, $h_y(\textbf{k}) = \xi k_x$ and $h_{\textrm{SOC}}(\textbf{k}) = \lambda_I k_y$, where $\lambda_I$ is the intrinsic spin-orbit coupling parameter. 

The presence of the SnSe layer breaks the $C_{2y}$ and $\mathcal{P}$ symmetries, but preserves the combined symmetry $C_{2y}C_{2z} = \mathcal{M}_y$. To zeroth order in $\textbf{k}$, the symmetry breaking proximity effect is described by 
\begin{eqnarray}
&H_{\textrm{prox}} = \Delta \sigma_z s_0 + \lambda_R \sigma_y s_y,
\label{eq3}
\end{eqnarray}
where $\lambda_R$ is the Rashba spin-orbit coupling and $\Delta$ is the sublattice symmetry breaking term originating from the proximity with the in-plane dipole field of SnSe. 

The sublattice symmetry breaking term, $\Delta$, counteracts the spin-orbit coupling gap in the spin subspace. Hence, the system undergoes a transition to a semimetallic state at a critical $|\Delta|$ value, henceforth refered to as $\Delta_c$, where electrons behave as pseudo-Weyl fermions with two-fold degenerate linear band crossings at two valleys located at $k_x = 0$ and $k_y = \eta \Lambda_y$, where $\eta = \pm 1$ is the valley index. The low-energy Hamiltonian for electrons at $\Delta = \Delta_c$ and $\lambda_R = 0$, without loss of generality, is~\cite{Snote}  
\begin{eqnarray}
H_{W}^{\eta s}(\textbf{q}) = (\Delta + s\eta \lambda_I\Lambda_y)\sigma_z + \hbar v_x q_x \sigma_y + \eta \hbar v_y q_y \sigma_x,
\label{eq_Hamiltonian}
\end{eqnarray}
where we have expanded $\textbf{k} = \textbf{q} + \eta (0, \Lambda_y)$ to linear order in $\textbf{q}$ with $|\textbf{q}|/\Lambda_y \ll 1$. Here, $s = \pm 1$ is the spin index and the velocities are $v_x \approx 2.38 \times 10^5$ m/s and $v_y \approx 4.44 \times 10^5$ m/s according to our first principles calculations~\cite{Snote}. The semimetallic bands are described by the Hamiltonian $H_{\textrm{PW}}^{\eta, s}(\textbf{q}) = v_x q_x \sigma_y + \eta v_y q_y \sigma_x$ with the constraint $\operatorname{sign}(\Delta_c) + s\eta = 0$. There is also a pair of gapped bands around each valley whose Hamiltonian is $H_{\textrm{G}}^{\eta, s}(\textbf{q}) = |2\Delta_c| \sigma_z + v_x q_x \sigma_y + \eta v_y q_y \sigma_x$ with the constraint $\operatorname{sign}(\Delta_c) - s\eta = 0$. The pseudo-Weyl state around the two valleys are time-reversal symmetric partners and, therefore, possess opposite spin polarizations.

We now describe the impact of the SnSe ferroelectric degree-of-freedom on the proximity-induced pseudo-Weyl state in the $\alpha$-Bi monolayer. At this level of description, the two SnSe in-plane polarization states are differentiated by the sign of the sublattice symmetry breaking parameter imprinted on the $\alpha$-Bi layer, i.e., $\Delta = +\Delta_c$ and $\Delta = - \Delta_c$ for opposite in-plane dipole fields. The constraint $\operatorname{sign}(\Delta_c) + s\eta = 0$ implies that in-plane polarization, valley and spin degrees-of-freedom are all coupled, such that the two ferroelectric states support opposite spin states at a given $\textbf{k}$. Hence, the spin states in the $\alpha$-Bi layer are coupled to the ferroelectric order parameter of SnSe through proximity effects.

Intuitively, the total spin-orbit field, $\textbf{B}_{\textrm{SO}}^{\textrm{total}}(\textbf{k})$, felt by $p_z$ electrons in $\alpha$-Bi is modified by the presence of the proximity-induced dipole field of the SnSe layer: $\textbf{B}_{\textrm{SO}}^{\textrm{total}}(\textbf{k}) \approx \textbf{B}_{\textrm{SO}}^{\textrm{CF}}(\textbf{k}) + \textbf{B}_{\textrm{SO}}^{\textrm{prox}}(\textbf{k})$, where $\textbf{B}_{\textrm{SO}}^{\textrm{CF}}(\textbf{k})$ is the intrinsic contribution due to the crystal field of monolayer $\alpha$-Bi and $\textbf{B}_{\textrm{SO}}^{\textrm{prox}}(\textbf{k})$ is due to proximity effects. To linear order in $\textbf{k}$, the inversion symmetry breaking proximity-induced contribution $\textbf{B}_{\textrm{SO}}^{\textrm{prox}}(\textbf{k}) \propto \textbf{P}\times \textbf{k}$ changes sign with the in-plane ferroelectric order parameter, $\textbf{P}$, of the SnSe layer. Therefore, reversal of $\textbf{P}$ causes spin states in $\alpha$-Bi layers to flip at a given $\textbf{k}$.

\begin{figure}[t]
\centerline{\includegraphics[width = \linewidth]{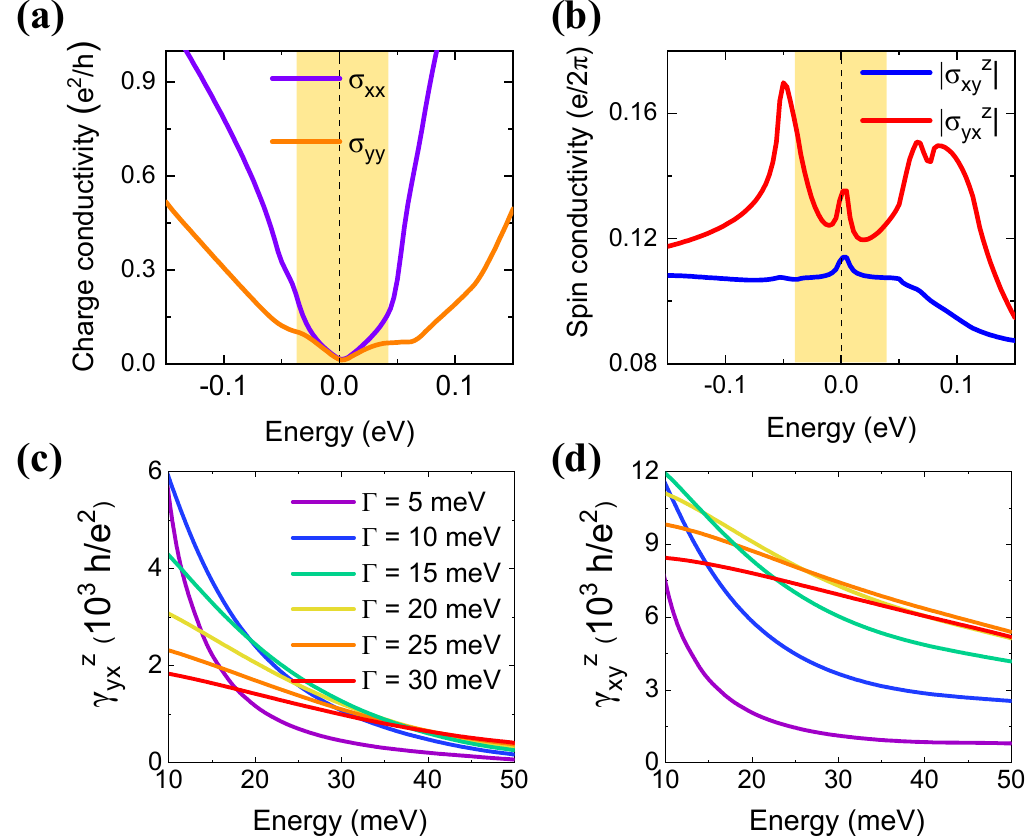}}
\caption{Spin-charge interconversion coefficients in $\alpha$-Bi/SnSe vdWHs. (a) charge and (b) spin Hall conductivities of  $\alpha$-Bi/SnSe vdWHs. Due to the anisotropic nature of electron states in $\alpha$-Bi, charge and spin conductivities along both in-plane directions are shown. (c) and (d) panels show the energy dependence of the spin-charge interconversion coefficients $\gamma_{\beta \alpha}^z = \sigma_{\beta \alpha}^z / \sigma_{\alpha \alpha}^2$ at several different spectral broadening for both in-plane directions, ($\alpha = x$ corresponding to the zigzag and $\alpha = y$ corresponding to the armchair).}
\label{Fig3}
\end{figure}

From the above analysis, $\alpha$-Bi/SnSe-type vdWHs are expected to host fully spin-polarized semimetallic states locked to the ferroelectric order parameter of SnSe in the limit $\Delta \rightarrow \Delta_c$. The fully spin polarized semimetallic states obtained from first principles are shown in Fig.~\ref{Fig2}(a). The pseudo-Weyl states described above are a good approximation for states within the highlighted yellow shaded energy window.

Having established the proximity-induced insulator-to-semimetal transition and spin-ferroelectricity locking, we now elaborate on the unique physical properties of $\alpha$-Bi/SnSe semimetals.

\emph{Berry Curvature Dipole.} \textemdash 
We begin by investigating the Berry curvature dipole (BCD)~\cite{PhysRevB.92.235447, PhysRevLett.115.216806} in $\alpha$-Bi/SnSe vdWHs. A previous work~\cite{Jin2021} revealed large and electrically-tunable BCD in $\alpha$-Bi monolayers. Because $\alpha$-Bi is a topological insulator, the large BCD is only accessible through doping, which can tune the Fermi level position towards the valence band~\cite{Li2021, Jin2021}. The insulator-to-semimetal transition induced by proximity-effects in $\alpha$-Bi/SnSe vdWHs, on the other hand, drastically modifies the low energy Bloch states of the $\alpha$-Bi layer and eliminates the necessity of hole doping due to the absence of an energy gap. These salient features indicate that proximity effects might induced large BCD at the Fermi level in $\alpha$-Bi/SnSe, which we investigate next.

The BCD is defined as
\begin{eqnarray}
D_{\alpha} = \sum_n\int \frac{d \textbf{k}}{(2\pi)^2} \frac{\partial f_{n\textbf{k}}}{ \partial k_{\alpha}} \Omega_{n\textbf{k}}^z,
\label{BCD}  
\end{eqnarray}
where $ f_{n\textbf{k}}$ is the Fermi-Dirac distribution function and $\Omega_{n\textbf{k}}^z$ is the z component of the Berry curvature. We obtain a Hamiltonian in a maximally localized basis by means of the \textrm{WANNIER90}~\cite{Mostofi2014, Pizzi2020} package and utilize it to compute the BCD and other related linear response quantities to be discussed in upcoming sections. Right panel of Fig.~\ref{Fig2}(a) displays the energy-dependent BCD for several spectral broadenings, $\Gamma$. Contrary to the freestanding $\alpha$-Bi case~\cite{Jin2021}, the $\alpha$-Bi/SnSe system displays remarkably large values at the immediate vicinity of the Fermi level, with opposite signs for valence and conduction states. The maximum value, ranging from $D_y \approx 1 - 4$ nm depending on the spectral broadening, is one order of magnitude larger than that of freestanding $\alpha$-Bi~\cite{Jin2021} and is mainly confined to the energy window where pseudo Weyl physics dominates the low energy description of electrons. Figures~\ref{Fig2}(b) and (c) show the k-space Berry curvature profile at the energies highlighted by the horizontal dashed lines in Fig.~\ref{Fig2}(a). Here, the berry curvature hotspots separated along the $k_y$ direction give rise to the BCD peaks seen in Fig.~\ref{Fig2}(a). At higher energies, the BCD drops substantially due to the spread of Berry curvature in momentum space, as shown in Fig.~\ref{Fig2}(c). This indicates that the highly confined berry curvature hotspots due to the pseudo Weyl states are responsible for the pronounced BCD features observed above, leading to a rapid change in nonlinear electromagnetic responses (such as nonlinear Hall) when the chemical potential is near the Weyl point.

It is worth noting that $D_x$ is constrained to vanish due to the mirror plane $\mathcal{M}_y$ and, hence, only the $D_y$ is finite. Furthermore, $D_y$ reverses sign upon switching the ferroelectric polarization of the SnSe layer. These features indicate that $\alpha$-Bi/SnSe vdWHs could enable unique non-linear responses, such as electrically controllable non-linear anomalous Hall effects~\cite{Jin2021}. This is further supported by reports on the electrically switchable ferroelectric polarization in SnSe(S) at room temperature~\cite{Higashitarumizu2020, Chang2020}. Next, we explore the charge-spin interconversion and its connection to the ferroelectricity of SnSe.

\emph{Spin-Charge Interconversion.} \textemdash
The coupling between spin states and ferroelectricity in $\alpha$-Bi/SnSe heterostructures offers great opportunities for the electrical manipulation of spin information in 2D systems via spin-to-charge and charge-to-spin conversion~\cite{Jafari2022}. The inverse spin Hall effect field $\textbf{E}^{ISHE}$ generated by an injected spin current $\textbf{Q}^z$, polarized out-of-plane, has components $E^{ISHE}_{\alpha} = (2e/\hbar) \gamma_{\beta\alpha}^z Q_{\beta}^z$, where $\gamma_{\beta\alpha}^{z} = \sigma_{\beta\alpha}^z/\sigma_{\alpha\alpha}^2$ is the spin-to-charge conversion efficiency with spin Hall and longitudinal charge conductivities given by $\sigma_{\beta\alpha}^z$ and $\sigma_{\alpha\alpha}$, respectively~\cite{Saitoh2006}.

 We compute the longitudinal Hall and spin Hall conductivities by means of linear response theory applied to the $\alpha$-Bi/SnSe wannier Hamiltonian. Figures~\ref{Fig3}(a) and (b) show the charge and spin conductivities. Due to the anisotropic nature of $\alpha$-Bi, $\sigma_{xx}$ and $\sigma_{yy}$ are quantitatively distinct, but with similar qualitative behavior; they exhibit a minimum at the fermi level due to the suppressed density of states of the semimetallic bands. The spin Hall conductivities, however, do not share such a minimum, as shown in Fig.~\ref{Fig3}(b); Instead, both components $\sigma_{xy}^z$ and $\sigma_{yx}^z$ assume finite values whose magnitudes are $\approx 0.12 (e/2\pi)$, for a finite spectral boradening of $\Gamma = 5$ meV according to our calculations. The anisotropy also impacts their relative magnitudes and qualitative dependencies on energy, where $\sigma_{xy}^z$ is almost insensitive to the fermi level while $\sigma_{yx}^z$ exhibits more salient features. These results indicate that $\alpha$-Bi/SnSe vdWHs are expected to support giant inverse spin Hall fields. 

The inverse spin Hall effect efficiencies are explicitly shown in Fig.~\ref{Fig3}(c) and (d) for several spectral broadening parameters $\Gamma$. We obtain giant efficiencies of thousand times the resistance quantum $h/e^2$, indicating the presence of large inverse spin Hall fields and responses. Although broadening and doping levels might impact the efficiency, all values obtained here are still roughly within the same order of magnitude. The dependence of the spin Hall conductivity with the substrate ferroelectric polarization is explicitly written as $\sigma_{xy}^z(\epsilon_F) = \operatorname{sign}(\Delta_c)\frac{e}{2\pi} g(\epsilon_F)$
where $g(\epsilon_F) = \int d^2\textbf{q} \sum_n f_{n\textbf{q}}^{+ \downarrow}(\epsilon_F)\Omega_{n, xy}^{+ \downarrow}(\textbf{q})/2\pi$ for one ferroelectric configuration with the integration performed over a single valley~\cite{Snote}, where dependence on the sign of $\Delta$ explicitly indicates ferroelectrically reversible direct and inverse spin Hall responses

In summary, we have shown that proximity-effects can be utilized to engineer 2D semimetallic states with spin-ferroelectricity locking, where ferroelectricity and spin states are confined to different layers, but are correlated by means of proximity effects. The semimetallic nature in these systems, supported by ARPES measurements, first principles calculations and minimal lattice models, were shown to support giant Berry curvature dipoles and spin Hall responses. Further, the spin-ferroelectricity locking also implies in non-volatile and electrically reversible linear and non-linear current responses in these systems, offering a perspective to enable highly-efficient control of unique quantum transport phenomenon in 2D materials.


\textit{Acknowledgments} 
T.L. and D.S. acknowledge partial support  from Office of Naval Research MURI grant N00014-23-1-2567. J.P.W. acknowledges the support of Robert F. Hartmann Endowed Chair Professorship. S.L. is supported by Basic Science Research Program through the National Research Foundation of Korea funded by the Ministry of Education (NRF-2021R1A6A3A14038837). The work at University of Missouri was supported by the U.S. Department of Energy, Office of Science, Office of Basic Energy Sciences, Division of Materials Science and Engineering, under Grant No. DE-SC0024294. G.B. acknowledges the support from Gordon and Betty Moore Foundation under Grant No. GBMF12247.  Q.L., R.G.M. and M.B. were supported by the U.S. Department of Energy, Office of Science, Basic Energy Sciences, Materials Science and Engineering Division.

\bibliographystyle{apsrev}
\bibliography{my.bib} 

\end{document}



\title{{\normalsize{Supplementary Information}} \\
Ferroelectric Semimetals with $\alpha$-Bi/SnSe van der Waals heterostructures and its Topological Currents}

\author{D. J. P. de Sousa}
\affiliation{Department of Electrical and Computer Engineering, University of Minnesota, Minneapolis, Minnesota 55455, USA}
\author{Seungjun Lee}
\affiliation{Department of Electrical and Computer Engineering, University of Minnesota, Minneapolis, Minnesota 55455, USA}
\author{Qiangsheng Lu}
\affiliation{Materials Science and Technology Division, Oak Ridge National Laboratory, Oak Ridge, Tennessee 37831, USA}
\author{Rob G. Moore}
\affiliation{Materials Science and Technology Division, Oak Ridge National Laboratory, Oak Ridge, Tennessee 37831, USA}
\author{Matthew Brahlek}
\affiliation{Materials Science and Technology Division, Oak Ridge National Laboratory, Oak Ridge, Tennessee 37831, USA}
\author{J-.P. Wang}
\affiliation{Department of Electrical and Computer Engineering, University of Minnesota, Minneapolis, Minnesota 55455, USA}
\affiliation{Department of Physics, University of Minnesota, Minneapolis, Minnesota 55455, USA}
\author{Guang Bian}\email{biang@missouri.edu}
\affiliation{Department of Physics and Astronomy, University of Missouri, Columbia, Missouri 65211, USA}
\author{Tony Low}\email{tlow@umn.edu}
\affiliation{Department of Electrical and Computer Engineering, University of Minnesota, Minneapolis, Minnesota 55455, USA}
\affiliation{Department of Physics, University of Minnesota, Minneapolis, Minnesota 55455, USA}

\date{\today}
\maketitle

\section*{Symmetries}\label{note0}
Let $D(g)$ be the representation for the symmetry operator $g$. Then, $g$ is a symmetry of the system if $D(g)H(\textbf{k})D^{-1}(g) = H(g\textbf{k})$. The symmetry action on the basis is
\begin{eqnarray}
& g \hat{c}_{\textbf{k}\alpha s}^{\dagger}g^{-1} = \displaystyle \sum_{s'}\hat{c}_{g\textbf{k},\alpha' s '}^{\dagger} [D(g)]_{\alpha' s', \alpha s}.
\label{symS1}
\end{eqnarray}

\begin{figure}[h]
\centerline{\includegraphics[scale = 0.8]{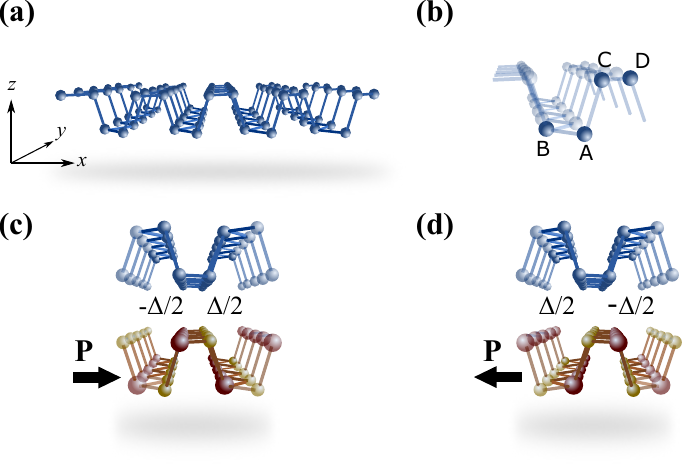}}
\caption{(a) Monolayer $\alpha$-Bi crystal structure. The sublattices of $\alpha$-Bi monolayer are highlighted in the right panel. (b) The basis consists of combinations of pairs of $p_z$ state amplitudes in sublattices A and D, and sublattices B and C. This is due to their symmetric relations such as in black-phosphorus monolayers~\cite{}. (c) and (d) illustrate the sublattice symmetry breaking due to proximity with the local polarization of SnSe. }
\label{FigS00}
\end{figure}

The symmetries of the low energy $p_z$ states of monolayer $\alpha$-Bi are two-fold rotations in relation to the z and x axis, $C_{2z}$ and $C_{2x}$ respectively, inversion $\mathcal{P}$ and time-reversal $\mathcal{T}$. Their representations are
\begin{eqnarray}
& D (C_{2x}) = i\sigma_x s_x, \nonumber \\
& D (C_{2z}) = -i\sigma_x s_z, \nonumber \\
& D (\mathcal{P}) = -i\sigma_x s_0, \nonumber \\
& D (\mathcal{T}) = -i\sigma_0 s_y \mathcal{K}, 
\label{symS2}
\end{eqnarray}
where $\mathcal{K}$ is the complex conjugation operator. The combined $D(C_{2z}C_{2y}) = -i\sigma_0 s_y$ symmetry corresponds to a $y$ mirror plane $\mathcal{M}_y$. Here, the $\sigma$ pauli matrices operator in orbital space and $s$ in spin space. The basis is $\Psi = [\phi_1 \ \ \phi_2]^{\dagger}$, where $\phi_j = [\phi_1^{\uparrow} \ \ \phi_1^{\downarrow}]^T$, with $\phi_1^{s} = \phi_{p_z}^{A,s} + \phi_{p_z}^{D,s} $ and $\phi_2^{s} = \phi_{p_z}^{B,s} + \phi_{p_z}^{C,s}$, where $\phi_{p_z}^{l,s}$ corresponds to the amplitude of finding electron in sublattice $l = A, B, C, D$ of the monolayer $\alpha$-Bi unit cell, which are identified in Fig.~\ref{FigS00}(b). 

Proximity with the in-plane polarization of SnSe breaks the symmetry between the sublattices A (D) and B (C). This is illustrated in Figs.~(\ref{FigS00}) (c) and (d) for the two ferroelectric configurations. Hence, proximity with SnSe breaks the two-fold rotation around the x axis, $C_{2x}$, and inversion, $\mathcal{P}$.  


\section*{Insulator-to-Semimetal transition}\label{note1}

We describe here the insulator-to-semimetal transition induced by proximity effects. To simplify our analysis, we assume $u_0 = \eta_x = \eta_y = \lambda_R = 0$, without loss of generality. Hence, the continuum Hamiltonian becomes
\begin{eqnarray}
&H(\textbf{k}) = h_x(\textbf{k}) \sigma_x s_0+ h_y(\textbf{k}) \sigma_y s_0 + h_{\textrm{SOC}}(\textbf{k}) \sigma_z s_z + \Delta \sigma_z s_0
\label{S1}
\end{eqnarray}
where $h_x(\textbf{k}) = \delta_0 + \gamma_x k_x^2 + \gamma_y k_y^2$, $h_y(\textbf{k}) = \xi k_x$ and $h_{\textrm{SOC}}(\textbf{k}) = \lambda_I k_y$, with $\lambda_I$ being the intrinsic spin-orbit coupling parameter that we assume to be positive, $\lambda_I > 0$.It is convenient to work in a basis containing symmetric and antisymmetric superpositions of all sublattice amplitudes. This is achieved by employing the unitary transformation $U = (\sigma_x + \sigma_z)s_0 /2$, which mixes the sublattice amplitudes as: $\bar{\Psi} = U\Psi = [\bar{\phi}_1 \ \ \bar{\phi}_2]^{\dagger}$, with $\bar{\phi}_1 = (\phi_A + \phi_D + \phi_B + \phi_C)/2$ and $\bar{\phi}_2 = (\phi_A + \phi_D - \phi_B - \phi_C)/2$. The transformed Hamiltonian, $\bar{H}(\textbf{k}) = U^{\dagger}H(\textbf{k}) U$, becomes,  
\begin{eqnarray}
& \bar{H}(\textbf{k}) =  h_x(\textbf{k}) \sigma_z s_0 - h_y(\textbf{k}) \sigma_y s_0 + \sigma_x(h_{\textrm{SOC}}(\textbf{k}) s_z + \Delta s_0).
\label{S2}
\end{eqnarray}

The gap closure takes place along the $k_x = 0$ region of momentum space. Hence, we solve for the eigenvalues of 
\begin{eqnarray}
& \bar{H}(k_y) = \displaystyle \left(
\begin{tabular}{cc}
   $(\delta_0 + \gamma_y k_y^2)s_0$  & $\Delta s_0 + \lambda_I k_y s_z$  \\
   $\Delta s_0 + \lambda_I k_y s_z$ & $-(\delta_0 + \gamma_y k_y^2)s_0$
\end{tabular}
\right).
\label{S3}
\end{eqnarray}
To this end, we explicitly write the Schr\"odinger equation, $\bar{H}(k_y)\bar{\Psi} = \epsilon(k_y)\bar{\Psi}$, in terms of its components
\begin{eqnarray}
\displaystyle (\delta_0 + \gamma_y k_y^2 - \epsilon)\bar{\phi}_1 + (\Delta s_0 + \lambda_I k_y s_z)\bar{\phi}_2 =&  0, \nonumber \\
\displaystyle (\delta_0 + \gamma_y k_y^2 + \epsilon)\bar{\phi}_2 - (\Delta s_0 + \lambda_I k_y s_z)\bar{\phi}_1 =&  0.
\label{S4}
\end{eqnarray}
The above pair of equations can be easily decoupled into
\begin{eqnarray}
\{[(\delta_0 + \gamma_y k_y^2)^2 - \epsilon^2]s_0 + (\Delta s_0 + \lambda_I k_y s_z)^2\}\bar{\phi}_1 = 0,
\label{S5}
\end{eqnarray}
which can be used to obtain the spin components of the symmetric combination of the sublattice amplitudes. The energy eigenvalues corresponding to the two distinct spin states are easily obtained by requiring the existence of non-trivial solutions for the components of $\bar{\phi}_1$. The results is 
\begin{eqnarray}
\epsilon_{ns}^{\eta}(k_y) = n\sqrt{(\delta_0 + \gamma_y k_y^2)^2 + \Delta^2 + \lambda_I^2 k_y^2 + 2s\lambda_I\Delta k_y},
\label{S6}
\end{eqnarray}
where $n = \pm1$ and $s = \pm 1$ are the band and spin indexes, respectively. Note that in the absence of spin-orbit coupling and sublattice symmetry breaking, i.e., $\lambda_I = \Delta = 0$, a gap closure takes place if $\operatorname{sign}(\delta_0\gamma_y) = -1$ at $\Lambda_y^{\eta} = \eta \sqrt{-\delta_0/\gamma_y}$, where $\eta = \pm$ being the valley index. The band crossings are protected by a mirror symmetry...distinct symmetry eigenvalues etc.

\begin{figure}[t]
\centerline{\includegraphics[scale = 0.5]{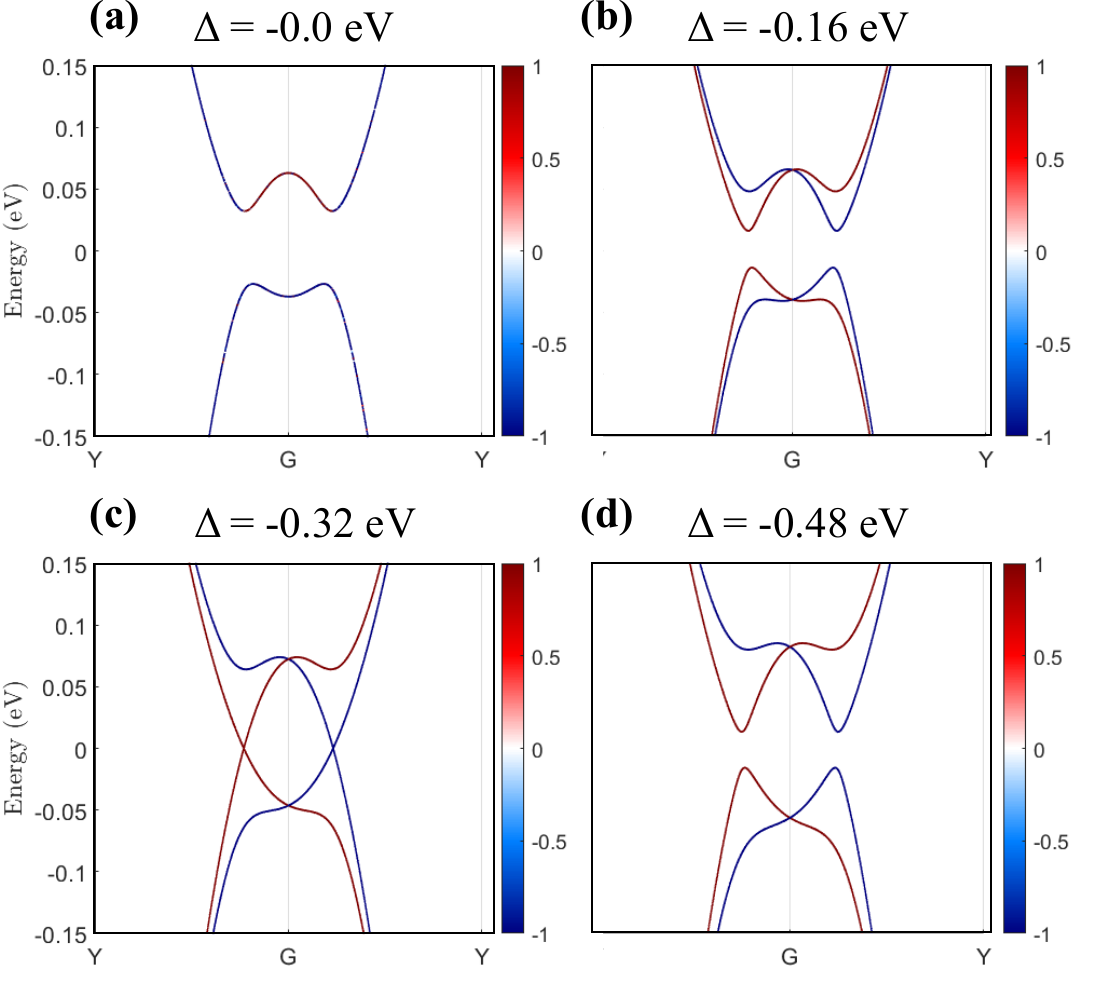}}
\caption{Insulator-to-semimetal transition; Evolution of the spin-resolved bands of the $p_z$ Hamiltonian of $\alpha$-Bi with the proximity-induced sublattice symmetry breaking parameter $\Delta$. We considered the parameter values $u_0 =  0$ eV, $\eta_x = 0.1$ eV \AA\ $^2$, $\eta_y = 0.5$ eV \AA\ $^2$, $\delta_0 = -0.05$ eV, $\gamma_x = 0.13$ eV \AA\ $^2$, $\gamma_y = 1.925$ eV \AA\ $^2$, $\lambda_I = -0.2$ eV \AA\ , $\lambda_R = 0$ eV , with (a) $\Delta = 0$ eV, (b) $\Delta = -0.016$  eV, (c) $\Delta = -0.032$ eV $ = \Delta_c$ and (d) $\Delta = -0.048$  eV.}
\label{FigS0}
\end{figure}

In the presence of intrinsic spin-orbit coupling and polarization-induced sublattice symmetry breaking terms, i.e., $\lambda_I \neq 0, \Delta \neq 0$, it is still possible to attain a gapless phase with band crossings at the valleys $\Lambda_y^{\eta}$. In this scenario, the energies become
\begin{eqnarray}
\epsilon_{ns}^{\eta}(k_y = \eta \Lambda_y) = n\sqrt{\Delta^2 + \lambda_I^2 \Lambda_y^2 + 2s\eta\lambda_I\Delta \Lambda_y},
\label{S7}
\end{eqnarray}
where $\Lambda_y = \sqrt{-\delta_0/\gamma_y}$, with an energy gap given by
\begin{eqnarray}
E_g = 2|\Delta + s\eta\lambda_I \Lambda_y|.
\label{S8}
\end{eqnarray}

Note that a gap closure can be attained for a particular choice of $\Delta$. We define the quantity $\Delta_c = \lambda_I \sqrt{-\delta_0 / \gamma_y}$, which will be very useful in the following. Hence, an insulator-to-semimetal transition takes place with $\Delta$. This is shown in Fig.~(\ref{FigS0}), where we have considered a lattice version of the symmetry derived Hamiltonian presented in the main text. As $\Delta$ is increased from $0$ eV to a finite value, the previously degenerated opposite spin bands split in momentum due to inversion symmetry breaking. The spin degeneracy breaking is accompanied by a decrease in the energy gap of opposite spin bands around  $\pm\Lambda_y$, which eventually vanishes at the critical value $\Delta_c$, as shown in Fig.~\ref{FigS0}(c). An energy gap reopens for $\Delta > \Delta_c$. 

Next, we examine two distinct situations: I) $\Delta > 0$ and II) $\Delta < 0$, corresponding to the two distinct polarization states. 

\subsection*{(I) The $\Delta > 0$ case}
The gap closes, i.e., $E_g = 0$, at the critical value $\Delta = \Delta_c$ whenever $\eta s = -1$. The first possibility, with $\eta = +1$ and $s = -1$, corresponds to a gap closure of the spin down bands at valley located at $k_y = +\Lambda_y$. The second possibility, with $\eta = -1$ and $s = +1$, corresponds to a gap closure of the spin up bands at the valley located at $k_y = -\Lambda_y$. Note that the spin down (up) bands remain gapped at $k_y = -\Lambda_y$ ($k_y = +\Lambda_y$). The gaps are $E_g = 4\lambda_I \sqrt{-\delta_0 / \gamma_y}$. Therefore, the system becomes semimetallic with two fully spin-polarized valleys separated along the $k_y$ direction. The opposite spin polarization of the two valleys is a manifestation of time-reversal symmetry. 

\subsection*{(II) The $\Delta < 0$ case}
A gap closure is attained at the critical value $\Delta = -\Delta_c$ whenever $\eta s = +1$. The first possibility, with $\eta = +1$ and $s = +1$, corresponds to a gap closure of the spin up bands at valley located at $k_y = +\Lambda_y$. The second possibility, with $\eta = -1$ and $s = -1$, corresponds to a gap closure of the spin down bands at the valley located at $k_y = -\Lambda_y$. The spin up (down) bands remain gapped at $k_y = -\Lambda_y$ ($k_y = +\Lambda_y$), with $E_g = 4\lambda_I \sqrt{-\delta_0 / \gamma_y}$. Hence, the system is a semimetal at $\Delta = -\Delta_c$ with two fully spin-polarized valleys separated along the $k_y$ direction. 

\begin{figure}[t]
\centerline{\includegraphics[scale = 0.5]{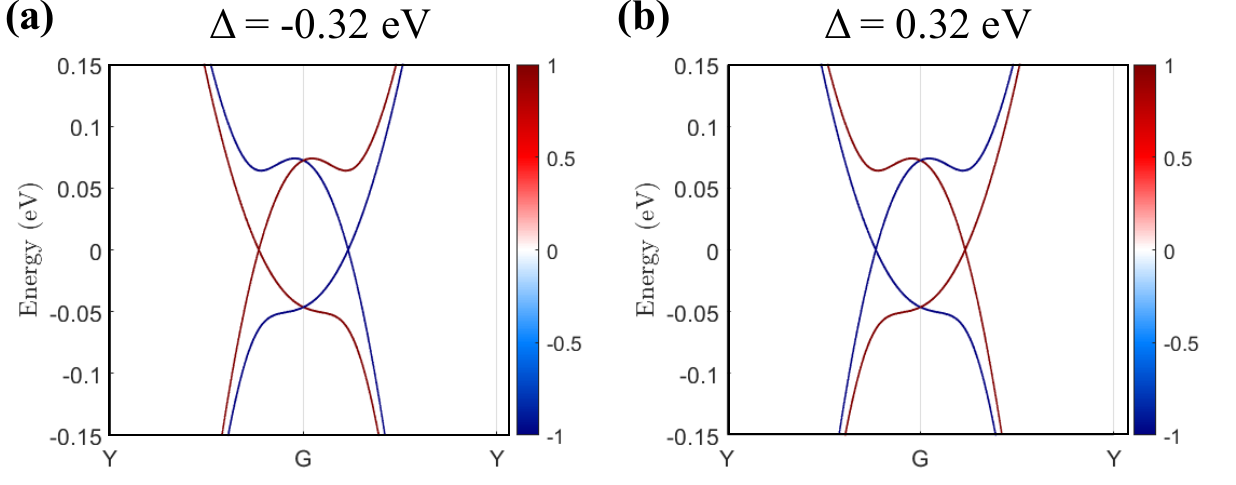}}
\caption{Locking between valley, spin and the sign of $\Delta$. Upon switching the ferroelectric polarization of SnSe layer, and consequently the sign of $\Delta$, the spin flavor in a given valley is reversed. We considered the parameter values $u_0 =  0$ eV, $\eta_x = 0.1$ eV \AA\ $^2$, $\eta_y = 0.5$ eV \AA\ $^2$, $\delta_0 = -0.05$ eV, $\gamma_x = 0.13$ eV \AA\ $^2$, $\gamma_y = 1.925$ eV \AA\ $^2$, $\lambda_I = -0.2$ eV \AA\ , $\lambda_R = 0$ eV , with (a) $\Delta = -0.032$ eV, (b) $\Delta = 0.032$  eV.}
\label{FigS2}
\end{figure}

The two situations described above are shown in Fig.~\ref{FigS2}. We conclude that the only effect of changing the sign of $\Delta$ when $|\Delta| = \Delta_c$ is to change the spin polarizations of the two valleys. Hence, the valley spin polarization, and consequently, the spin/Valley Hall effect are coupled to the polarization states of the SnSe monolayer underneath $\alpha$-Bi. 


\section*{Effective pseudo-Weyl Hamiltonian}\label{note2}

From the previous analysis, one should expect Weyl-like Hamiltonians describing the electron states around each valley. Because the crossing bands are fully spin polarized, we proceed by extracting the Hamiltonians for each spin sector. To this end, we start by rewriting Hamiltonian~(\ref{S1}) in the ``spin-polarized" basis $\Tilde{\Psi} = [\phi_1^{\uparrow} \ \ \phi_2^{\uparrow} \ \ \phi_1^{\downarrow} \ \ \phi_2^{\downarrow}]^T$. This is achieved through the unitary transformation
\begin{eqnarray}
& \Tilde{U} = \displaystyle \left(
\begin{tabular}{cccc}
   1 & 0 & 0 & 0 \\
   0 & 0 & 1 & 0 \\
   0 & 1 & 0 & 0 \\
   0 & 0 & 0 & 1 
\end{tabular}
\right).
\label{S9}
\end{eqnarray}

The transformed Hamiltonian, $\Tilde{H}(\textbf{k}) = \Tilde{U}^{\dagger} H(\textbf{k})\Tilde{U}$, is block diagonal in spin-space, $\Tilde{H}(\textbf{k}) = \operatorname{diag}[H^{\uparrow}(\textbf{k}) \ \ H^{\downarrow}(\textbf{k})]$ , with
\begin{eqnarray}
& H^{s}(\textbf{k}) = \displaystyle \left(
\begin{tabular}{cc}
   $\Delta + s h_{\textrm{SOC}}(\textbf{k})$ & $h_x(\textbf{k}) - ih_y(\textbf{k})$ \\
   $h_x(\textbf{k}) + ih_y(\textbf{k})$ & $-\Delta - s h_{\textrm{SOC}}(\textbf{k})$
\end{tabular}
\right),
\label{S10}
\end{eqnarray}
where $s = \pm 1$ refers to the two spin states. Note that Hamiltonian~(\ref{S10}) assumes the compact form $H^s(\textbf{k}) = \textbf{h}^s(\textbf{k}) \cdot \boldsymbol{\sigma}$, where $\textbf{h}^s(\textbf{k}) = (h_x^s(\textbf{k}), h_y^s(\textbf{k}), h_z^s(\textbf{k}))$ with $h_x^s(\textbf{k}) = h_x(\textbf{k})$, $h_y^s(\textbf{k}) = h_y(\textbf{k})$ and $h_z^s(\textbf{k}) = \Delta + s h_{\textrm{SOC}}(\textbf{k})$. Later, we are going to utilize this fact to investigate the spin Hall response. If $h_0(\textbf{k}) \neq 0$, we obtain $H^s(\textbf{k}) = h_0(\textbf{k})\sigma_0 + \textbf{h}^s(\textbf{k}) \cdot \boldsymbol{\sigma}$.

Expanding Eq.~(\ref{S10}) around the valley $\eta \Lambda_y$, i.e, assuming $\textbf{k} = \textbf{q} + \eta (0, \Lambda_y)$ to linear order in $\textbf{q}$ with $|\textbf{q}|/\Lambda_y \ll 1$, we find
\begin{eqnarray}
H_W^{\eta s}(\textbf{q}) = (\Delta + s\eta \lambda_I\Lambda_y)\sigma_z + \hbar v_x q_x \sigma_y + \eta \hbar v_y q_y \sigma_x,
\label{S11}
\end{eqnarray}
which is an anistropic version of a pseudo-Weyl Hamitonian, with pseudospin (sublattice) degree of freedom replacing the real spin, with a spin-valley dependent mass term. The velocities are $v_x = \xi /\hbar$ and $v_y = 2\sqrt{-\delta_0 \gamma_y}/\hbar$. By fitting the first principles band structure, we obtain that the velocities are $v_x \approx 2.38 \times 10^5$ m/s and $v_y \approx 4.44 \times 10^5$ m/s, encoding information about the structural anisotropy.


\section*{Topological Phase Transition Induced by in-Plane Ferroelectricity}\label{note_Pfaffian}

We present here the details of the quantum spin Hall state transitions induced by fluctuations in $\Delta$ away from $\Delta_c$. The antisymmetric matrix $\langle u_{n}(\textbf{k})|\mathcal{T}|u_m(\textbf{k})\rangle$ for the occupied states $|u_{n,m}(\textbf{k})\rangle$ has a single element $\langle u_{1}(\textbf{k})|\mathcal{T}|u_2(\textbf{k})\rangle$ which corresponds to its Pfaffian in our four-bands model. Transitions between topologically trivial and non-trivial insulating phases are accompanied by the absence or presence, respectively, of zeros of the Pfaffian in the half Brillouin zone~\cite{}. Thus, we proceed by explicitly calculating the Pfaffian $P(\textbf{k}) = \langle u_{1}(\textbf{k})|\mathcal{T}|u_2(\textbf{k})\rangle$ over half Brillouin zone.

The normalized eigenstates of $H^s(\textbf{k}) = h_0(\textbf{k})\sigma_0 + \textbf{h}^s(\textbf{k}) \cdot \boldsymbol{\sigma}$ are
\begin{eqnarray}
    |u^s_{n}(\textbf{k})\rangle = \displaystyle \frac{1}{\sqrt{2h_s(\textbf{k})}}\left(
    \begin{tabular}{c}
         $\sqrt{h_s(\textbf{k}) + n m_s(\textbf{k})} e^{-i\phi(\textbf{k})/2}$ \\
         $n\sqrt{h_s(\textbf{k}) - n m_s(\textbf{k})} e^{i\phi(\textbf{k})/2}$ 
    \end{tabular}
    \right),
    \label{Pfaffian_eq1}
\end{eqnarray}
where $n = \pm 1$ is the band index, $s = \uparrow, \downarrow$ is the spin index, $h_s(\textbf{k}) = \sqrt{h_x(\textbf{k})^2 + h_y(\textbf{k})^2 + m_s(\textbf{k})^2}$, $m_s(\textbf{k}) = \Delta + s h_{SOC}(\textbf{k})$ and $\phi(\textbf{k}) = \arctan(h_y(\textbf{k})/h_x(\textbf{k}))$. The associated energies are $\epsilon_n^s(\textbf{k}) = h_0(\textbf{k}) + n h_s(\textbf{k})$. The states with the lowest energies $\epsilon_1 = \epsilon_{-}^{\uparrow}$ and $\epsilon_2 = \epsilon_{-}^{\downarrow}$ are
\begin{eqnarray}
    |u_{1}(\textbf{k})\rangle = \displaystyle \left(
    \begin{tabular}{c}
         $\mathcal{A}_{-}^{\uparrow}(\textbf{k}) e^{-i\phi(\textbf{k})/2}$ \\
         -$\mathcal{A}_{+}^{\uparrow}(\textbf{k}) e^{i\phi(\textbf{k})/2}$ \\
         0 \\
         0
    \end{tabular}
    \right), \ \ 
    |u_{2}(\textbf{k})\rangle = \displaystyle \left(
    \begin{tabular}{c}
         0 \\
         0 \\
          $\mathcal{A}_{-}^{\downarrow}(\textbf{k}) e^{-i\phi(\textbf{k})/2}$ \\
         -$\mathcal{A}_{+}^{\downarrow}(\textbf{k}) e^{i\phi(\textbf{k})/2}$
    \end{tabular}
    \right), 
    \label{Pfaffian_eq2}
\end{eqnarray}
respectively. Here, $\mathcal{A}_{n}^s(\textbf{k}) = \sqrt{[h_{s}(\textbf{k}) + n m_{s}(\textbf{k})]/2h_{s}(\textbf{k})}$. The time-reversal operator in this basis is
\begin{eqnarray}
& \mathcal{T} = \displaystyle \left(
\begin{tabular}{cccc}
   0 & 0 & -1 & 0 \\
   0 & 0 & 0 & -1 \\
   1 & 0 & 0 & 0 \\
   0 & 1 & 0 & 0 
\end{tabular}
\right)\mathcal{K}.
\label{Pfaffian_eq3}
\end{eqnarray}

A direct computation gives
\begin{eqnarray}
P(\textbf{k}) = -\mathcal{A}_{-}^{\uparrow}(\textbf{k})\mathcal{A}_{-}^{\downarrow}(\textbf{k}) e^{i\phi(\textbf{k})} -\mathcal{A}_{+}^{\uparrow}(\textbf{k})\mathcal{A}_{+}^{\downarrow}(\textbf{k}) e^{-i\phi(\textbf{k})}.
    \label{Pfaffian_eq4}
\end{eqnarray}
Figure~\ref{Pfaffian_eq4} shows the evolution of the Pfaffian with $\Delta$ over the half Brillouin zone $k_y > 0$. For $\Delta = 0$ eV, the Pfaffian zeros occur along a ring aroung the $\Gamma$ point. At $\Delta = -16$ meV, a single Pfaffian zero occur at $k_x = 0$, indicating that the system is a quantum spin Hall insulator; Time-reversal symmetry topologically protects such zero from disappearing. For $\Delta = -48$ meV, there are no Pfaffian zeros and the system is in a trivial insulating phase. The transition to trivial insulator occurs at $\Delta = \Delta_c$ through the pseudo-Weyl phase, as described in previous sections. 

\begin{figure}[t]
\centerline{\includegraphics[width = \linewidth]{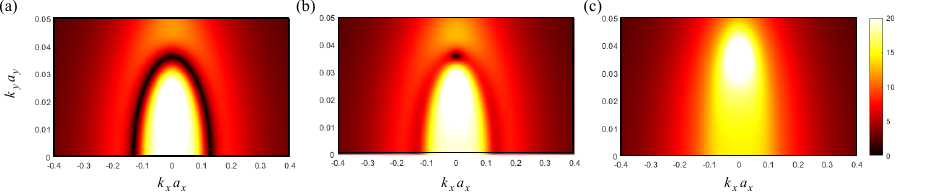}}
\caption{Evolution of the Pfaffian of $\langle u_n(\textbf{k})| \mathcal{T} | u_m(\textbf{k})\rangle$ over half of Brillouin zone for (a) $\Delta = 0$ meV, (b) $\Delta = -16$ meV and (c) $\Delta = -48$ meV.}
\label{Pfaffian_map}
\end{figure}

The presence or absence of a Pfaffian zero is associated with a change of the sign of the Dirac mass term. To show this is in fact the case, we first assume $\Delta \neq 0$. Because the Pfaffian zero is confined at $k_x = 0$ in this case, $h_y(\textbf{k}) = 0$ and $h_x(\textbf{k}) = \delta_0 + \gamma_y k_y^2$. Further simplification can be attained by focusing on $k_y \approx \Lambda_y$, which enables us to assume $h_x(\textbf{k}) \approx 0$. Within these assumptions $\mathcal{A}_{n}^s(\textbf{k}) \approx \sqrt{[|m_{s}(\textbf{k})| + n m_{s}(\textbf{k})]/2|m_{s}(\textbf{k})|}$. Note that $|z| \pm z = [1 \pm \operatorname{sign}(z)]|z|$ for arbitrary $z$. Therefore,  $\mathcal{A}_{n}^s\approx \sqrt{[1 + n \operatorname{sign}(m_s)]/2}$. If $m_{\uparrow} < 0$, as in our case, $\mathcal{A}_{+}^{\uparrow} = 0$ and 
\begin{eqnarray}
|P(\textbf{k})| = \sqrt{(1 - \operatorname{sign}(m_{\downarrow}))/2},
    \label{Pfaffian_eq4}
\end{eqnarray}
showing that a mass sign change results in transitions between finite and vanishing $|P(\textbf{k})|$. 

We conclude that fluctuations of $\Delta$ around the critical value $\Delta_c$ will transition the system to topologically trivial or non-trivial insulating states. Because, the polarization magnitude in two-dimensional SnSe-type ferroelectrics can be tuned by perpendicular electric fields or strain~\cite{}, we argue that the topological phase transitions can be fully controlled in $\alpha$-Bi/SnSe van der Waals heterostructures.


\section*{Topological surface states}\label{note_surface_states}

In this section, we study the $\alpha$-Bi/SnSe bilayer surface states and their dependence on the ferroelectric polarization of SnSe from the perspective of the lattice model. Free standing $\alpha$-Bi is a time-reversal symmetric topological insulator and, therefore, possesses topologically robust helical surface states whose energies lie within its bulk band gap. From the previous section, we anticipate an unique interplay between topological surface states and ferroelectricity across the insulator-to-semimetal transition, which raises the question about how the helical surface states evolve.

\begin{figure}[t]
\centerline{\includegraphics[width = \linewidth]{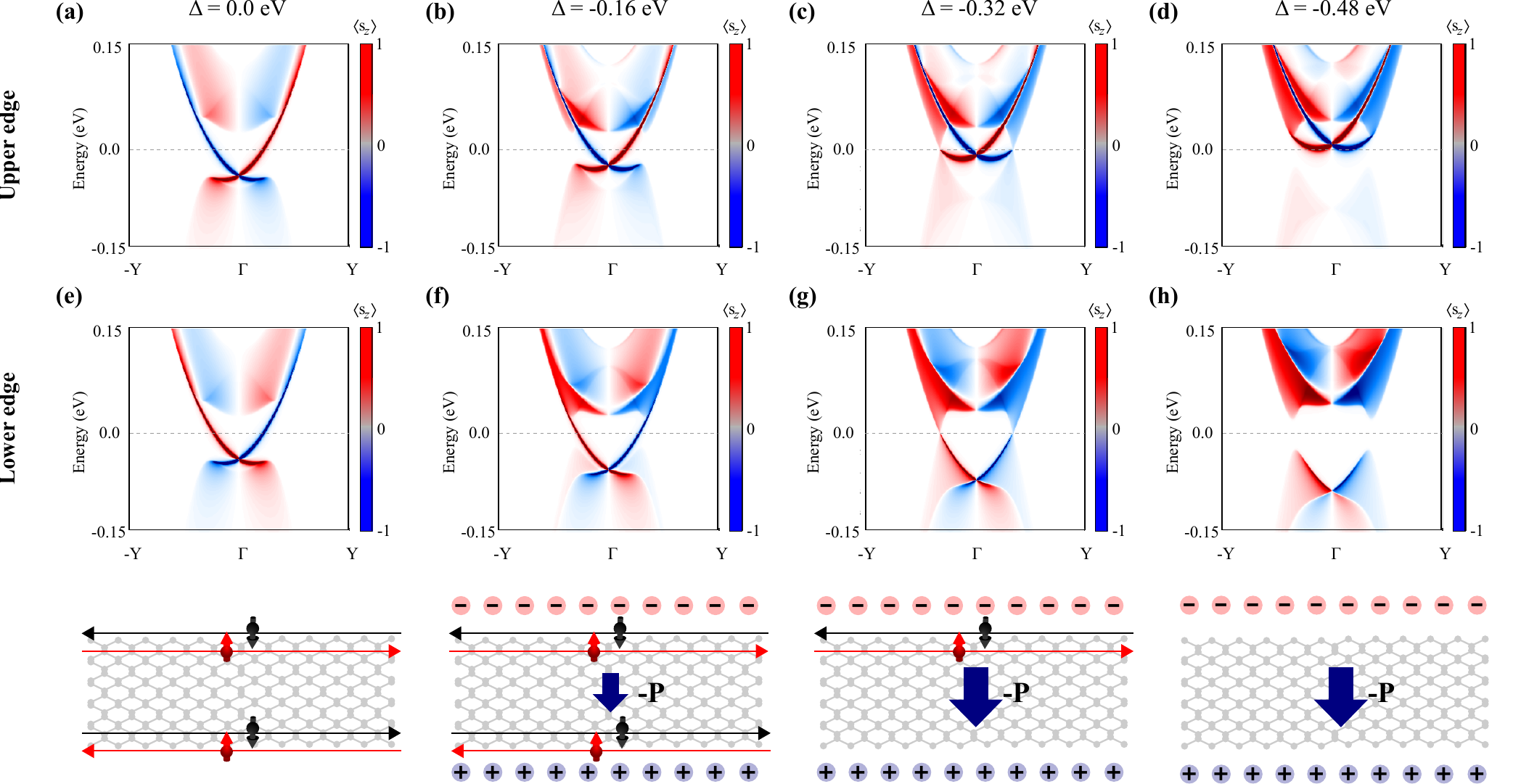}}
\caption{Evolution of topological surface states along the zigzag edges with the polarization-induced symmetry breaking parameters $\Delta$ across the insulator-to-semimetal transition. Local spin density of states intensity map for the upper edge in panels (a)-(d) and for the lower edge in panels (e)-(h). The bottom sketches summarize the information about the evolution of helical surface states with the SnSe polarization.}
\label{FigS_surface_state_1}
\end{figure}

To investigate the evolution of the topological surface modes, we can discretize the bulk Hamiltonian along the $x$ or $y$ direction, for armchair and zigzag edges. This gives 
\begin{eqnarray}
\mathcal{\hat{H}} = \sum_{j} [H_S(\textbf{k}_{||})c_{\textbf{k}_{||}, j}^{\dagger}c_{\textbf{k}_{||}, j} + \hat{W}_{\rm{c}}(\textbf{k}_{||})c_{\textbf{k}_{||}, j+1}^{\dagger}c_{\textbf{k}_{||}, j} + \hat{W}_{\rm{c}}^{\dagger}(\textbf{k}_{||})c_{\textbf{k}_{||}, j-1}^{\dagger}c_{\textbf{k}_{||}, j}],
\end{eqnarray}
from where the ``onsite" and ``hopping" Hamiltonians for the discretized system can be obtained as $H_S$ and $W_c$, respectively. In addition, $\textbf{k}_{||} = k_x$ or $\textbf{k}_{||} = k_y$ is momentum along the direction where the system is still translationally periodic, $j$ is the principal layer (PL) index and $c_{\textbf{k}_{||}, j}$ ($c_{\textbf{k}_{||}, j}^{\dagger}$) annihilates (creates) a particle with momentum $\textbf{k}_{||}$ at the $j$-th PL. A semi-infinite lead is described by a tridiagonal block matrix Hamiltonian, given by
\begin{eqnarray}
H = \left(
\begin{array}{cccc}
\ddots & \ddots &  &  \\
\ddots & H_{\rm{S}} &\hat{W}_{\rm{c}} &  \\
 & \hat{W}_{\rm{c}}^{\dagger} & H_{\rm{S}} & \hat{W}_{\rm{c}} \\
 &  & \hat{W}_{\rm{c}}^{\dagger} & H_{\rm{S}} \\
\end{array}
\right),
\label{eq1}
\end{eqnarray}
The surface spin density of states $S_z(\textbf{k}_{||}, \epsilon)$, is
\begin{eqnarray}
S_{\alpha}(\textbf{k}_{||}, \epsilon) = -\displaystyle \frac{1}{\pi}\textrm{Im}\textrm{Tr} [ g_{\textrm{S}}(\textbf{k}_{||}, \epsilon + i\eta)\sigma_0 s_{\alpha}], \label{eq2b}  
\end{eqnarray}
where $\eta = 10^{-3}$ eV is the broadening parameter assumed in all calculations. 

Figure~\ref{FigS_surface_state_1} shows the evolution of the helical surface states, along both zigzag edges referred to as upper and lower edges, with the polarization-induced symmetry breaking parameters $\Delta$ across the insulator-to-semimetal transition. At $\Delta = 0$ eV, upper and lower edge states are related by a $\mathcal{M}_x$ mirror plane, as seen from Figs.~\ref{FigS_surface_state_1}(a) and (e). The system is in a quantum spin Hall state with dissipationless couterpropagating spins, as shown in the corresponding skectch. The $\Delta = 0$ eV corresponds to the free standing $\alpha$-Bi case. 

At $\Delta = -0.016$ eV, the $\mathcal{M}_x$ symmetry is broken due to the finite bulk polarization. Hence, upper and lower edge states cannot be related by such a symmetry and their helical surface state dispersion now differ qualitatively, as shown in Figs.~\ref{FigS_surface_state_1}(b) and (f). We interpret such an asymmetric feature as being a consequence of the development of surface charges on the zigzag edges due to the finite bulk polarization. Here, the surface charges can only perturb the surface states, but cannot destroy them since a topological phase transition has not taken place. 

\begin{figure}[t]
\centerline{\includegraphics[width = \linewidth]{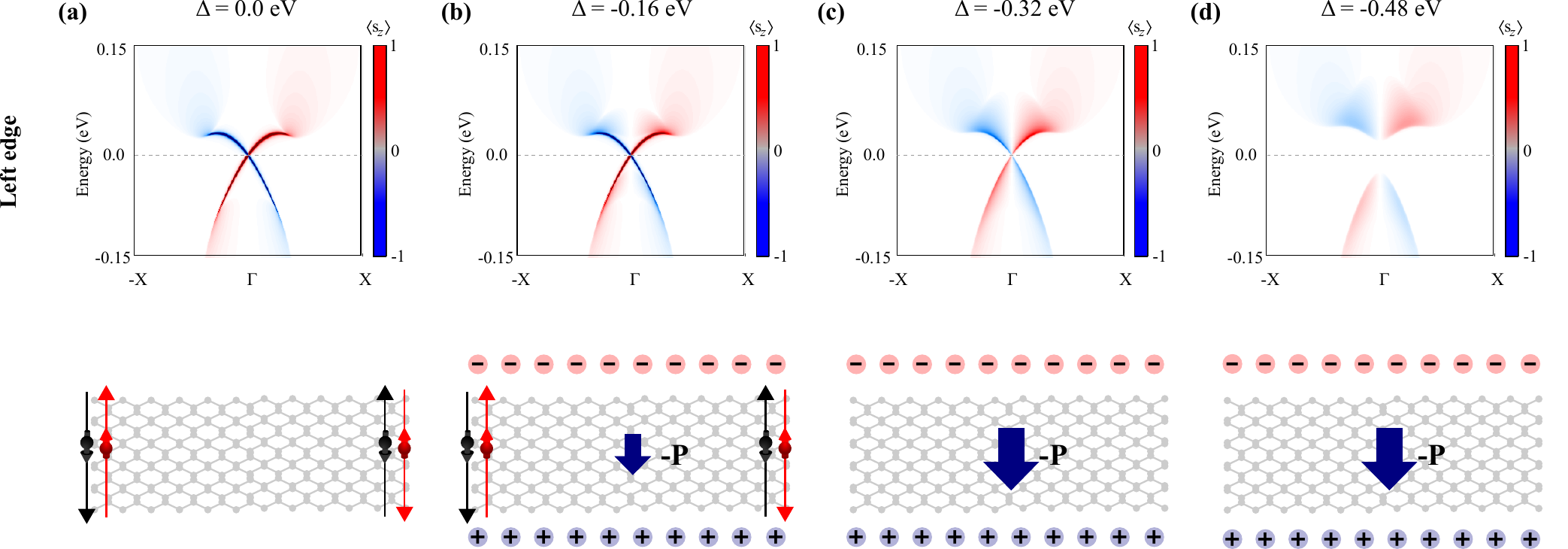}}
\caption{Evolution of topological surface states along the armchair edges with the polarization-induced symmetry breaking parameters $\Delta$ across the insulator-to-semimetal transition. Local spin density of states intensity map for the upper and bottom edges in panels (a)-(d). In this case, the two edges are related by the mirror symmetry $\mathcal{M}_y$. The bottom sketches summarize the information about the evolution of helical surface states with the SnSe polarization.}
\label{FigS_surface_state_2}
\end{figure}

At $\Delta = - 0.032$ eV, the critical transition value $\Delta_c$, a gap closure is attained and the system transitions to a semimetallic phase. Note that the upper zigzag surface still host helical edge states, Figs.~\ref{FigS_surface_state_1}(c), while the lower zigzag edge does not, Figs.~\ref{FigS_surface_state_1}(g). Here, the surface charges on the lower edge are enough to dissolve the edge states into the bulk, which is accompanied by the merging of the helical modes to the bulk bands, as shown Figs.~\ref{FigS_surface_state_1}(g). 

At $\Delta = -0.048$ eV, or higher values, the energy gap reopens again and the system is now topologically trivial with no helical states. The edge modes of the upper zigzag edge remain attached to conduction bands while those of the lower edge have been successfully become bulk valence states.

Figure~\ref{FigS_surface_state_2} shows the evolution of the helical surface states, along both armchair edges, with the polarization-induced symmetry breaking parameters $\Delta$ across the insulator-to-semimetal transition. Helical surface states are revealed at $\Delta = 0$ eV, consistent with the quantum spin Hall insulating state of free standing $\alpha$-Bi. At $\Delta \neq 0$ eV, we observe that the surface states of both armchair edges are identically the same, irrespective of the magnitude of $\Delta$. This is due to the fact that the polarization does not break the mirror plane $\mathcal{M}_y$ and, hence, the two armchair edges modes are related to each other. We only plot the local spin density of states for one of the armchair edges in Figure~\ref{FigS_surface_state_2}, which is referred to as the left edge and is illustrated in the corresponding sketch. As $\Delta$ is increased in magnitude across the transition point, the both edge modes are dissolved into the bulk.  

Note that the situation here is quite distinct from the zigzag edge case, where helical surface state still survive in one of the edges (upper). The interplay between polarization and topological surface states enables one to selectively control in which edge the helical states propagate by means of electric field switching.


\section*{Berry curvature and spin Hall Response}\label{note3}

Here, we examine the spin Hall effect and its dependence on the in-plane polarization. Because the spin Hamiltonians are decoupled, it is also possible to write the velocity of a state $n$ at valley $\eta$ for a given spin $s$. To linear order in electric field, $\textbf{E}$, the velocity is
\begin{eqnarray}
\textbf{v}_n^{\eta s}(\textbf{k}) = \displaystyle \frac{1}{\hbar}\frac{\partial \epsilon_{ns}^{\eta}(\textbf{k})}{\partial \textbf{k}} - \frac{e}{\hbar} \textbf{E} \times \boldsymbol{\Omega}_n^{\eta s}(\textbf{k}),
\label{S12}
\end{eqnarray}
where ${\Omega}_n^{\eta s}(\textbf{k})$ is the Berry curvature. The components of the Berry curvature vector appearing in the velocity are related to the components of the Berry curvature tensor, $\Omega_{n, \mu \nu}^{\eta s}(\textbf{k})$, through $\Omega_{n, \mu \nu}^{\eta s}(\textbf{k}) = \epsilon_{\mu \nu \rho} \Omega_{n, \rho}^{\eta s}(\textbf{k})$, where $\epsilon_{\mu \nu \rho}$ are the components of the Levi-Civit\`a antisymmetric tensor. 

Equation~(\ref{S11}) can be rewritten as general two-level system Hamiltonian, i.e., $H_W^{\eta s}(\textbf{q}) = \textbf{h}^{\eta s}(\textbf{q}) \cdot \boldsymbol{\sigma}$, for which the Berry curvature tensor components are most generally given by
\begin{eqnarray}
\Omega_{n, \mu \nu}^{\eta s}(\textbf{q}) = \displaystyle \frac{n}{2 [h^{\eta s}(\textbf{q}))]^3} \sum_{klm}\epsilon_{klm} h^{\eta s}_{k}(\textbf{q}) \frac{\partial h^{\eta s}_{l}(\textbf{q})}{\partial q_{\mu}} \frac{\partial h^{\eta s}_{m}(\textbf{q})}{\partial q_{\nu}},
\label{S13}
\end{eqnarray}
where $h^{\eta s}(\textbf{q}) = \sqrt{h_x^{\eta s}(\textbf{q})^2 + h_y^{\eta s}(\textbf{q})^2 + h_z^{\eta s}(\textbf{q})^2}$. The $xy$ component derived from Hamiltonian~(\ref{S11}) and Eq.~(\ref{S13}) is
\begin{eqnarray}
\Omega_{n, xy}^{\eta s}(\textbf{q}) = \displaystyle -n\frac{v_x v_y}{2}\frac{\eta \Delta + s\lambda_I\Lambda_y}{[(v_x q_x)^2 + (v_y q_y)^2 + (\eta \Delta + s \lambda_I \Lambda_y)^2]^{3/2}}, 
\label{S14}
\end{eqnarray}
which is fully time-reversal symmetric
\begin{eqnarray}
\Omega_{n, xy}^{\eta s}(\textbf{q}) = -\Omega_{n, xy}^{(-\eta) (-s)}(-\textbf{q}), 
\label{S15}
\end{eqnarray}
as one should expect. The explicit expression of the Berry curvature tensor for the two polarization states are given below
\begin{eqnarray}
\Omega_{n, xy}^{\eta s}(\textbf{q}) = \left\{ 
\begin{tabular}{cc}
    $\displaystyle  -n\frac{v_xv_y \Delta_c}{2}\frac{s + \eta}{[(v_x q_x)^2 + (v_y q_y)^2 + (\Delta_c)^2(s + \eta)^2]^{3/2}}$, & if $\Delta = +\Delta_c$ \\
     $\displaystyle  n\frac{v_xv_y \Delta_c}{2}\frac{s - \eta}{[(v_x q_x)^2 + (v_y q_y)^2 + (\Delta_c)^2(s - \eta)^2]^{3/2}}$, & if $\Delta = -\Delta_c$ 
\end{tabular}
\right., 
\label{S16}
\end{eqnarray}
from which we conclude that $\Omega_{n, xy}^{-\uparrow}(\textbf{q}\neq \textbf{0}) = \Omega_{n, xy}^{+\downarrow}(\textbf{q}\neq \textbf{0}) = 0$ for $\Delta =+\Delta_c$, whereas $\Omega_{n, xy}^{+\uparrow}(\textbf{q}\neq \textbf{0}) = \Omega_{n, xy}^{-\downarrow}(\textbf{q}\neq \textbf{0}) = 0$ for $\Delta =-\Delta_c$, and singular at $\textbf{q} = \textbf{0}$. Note that these Berry curvature components correspond to the semimetallic bands at the two valleys. The singular behavior is analogous to that in monolayer graphene, but with a two-fold degenerate band crossing in the $\alpha$-Bi/SnSe case.

\begin{figure}[t]
\centerline{\includegraphics[scale = 0.45]{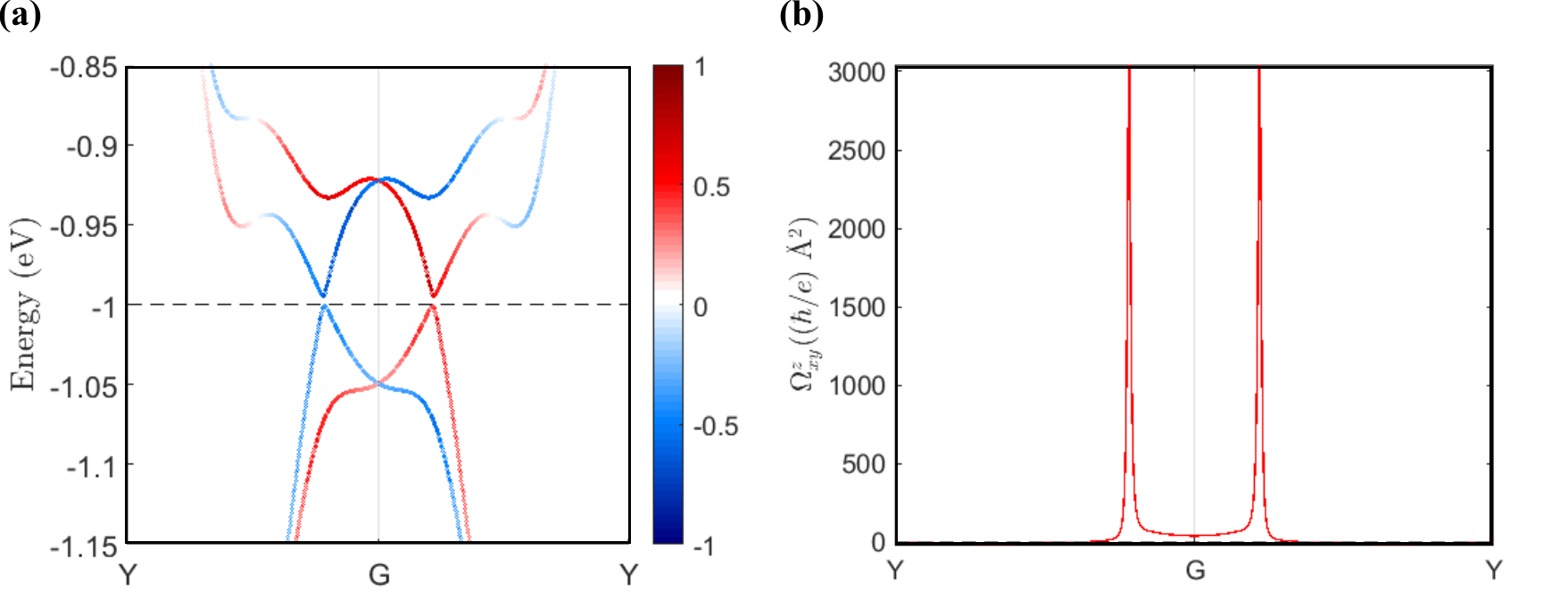}}
\caption{Fermi level Spin Berry curvature of $\alpha$-Bi/SnSe obtained from first principles. The two peaks around the two valleys are well described by Eq.~(\ref{S14}), where the linear pseudo-Weyl approximation is valid.}
\label{FigS3}
\end{figure}

The non-singular Berry curvature components correspond to the spin-polarized gapped bands. Their magnitudes are 
\begin{eqnarray}
\Omega_{n, xy}^{+ \uparrow}(\textbf{q}) = -\Omega_{n, xy}^{- \downarrow}(\textbf{q}) = 
    \displaystyle  -n\frac{v_x v_y\Delta_c}{[(v_x q_x)^2 + (v_y q_y)^2 + (2\Delta_c)^2]^{3/2}}, \ \ \Delta = +\Delta_c 
    \label{S17}
\end{eqnarray}
and
\begin{eqnarray}
\Omega_{n, xy}^{+ \downarrow}(\textbf{q}) = -\Omega_{n, xy}^{- \uparrow}(\textbf{q}) = 
    \displaystyle  -n\frac{v_x v_y\Delta_c}{[(v_x q_x)^2 + (v_y q_y)^2 + (2\Delta_c)^2]^{3/2}}, \ \ \Delta = -\Delta_c .
    \label{S18}
\end{eqnarray}

Thus, the total Berry curvature at a given valley is the sum of the Berry curvature of semimetallic and insulating bands. The contributions from the semimetallic bands is singular, whereas the contributions from the insulating bands is finite for a larger range of momenta around the valley.

Our first principles calculations, however, reveals a tiny energy gap, $\delta_g \approx 6.2$ meV, in the lowest energy bands of $\alpha$-Bi/SnSe, implying that the induced symmetry breaking polarization due to the proximity with SnSe, $\Delta_{\textrm{SnSe}}$, slightly deviates from the critical value, estimated to be $\Delta_c \approx 33.8$ meV. Hence, we identify $\delta_g = 4|\Delta_{\textrm{SnSe}} - \Delta_c|$, where $\delta_g/4\Delta_c \approx 0.046 \ll 1$ is the gap ratio between lower and higher energy bands in $\alpha$-Bi/SnSe. As a consequences, the Berry curvature components of the bands with lower energy is non-singular and finite over a certain range in momentum space.  Its magnitude is much larger than that of the gapped bands. Therefore, when analysing the Berry curvature components of the lower energy bands, we will assume a small energy gap. The perfect semimetal limit can also be obtained by leaving $\delta_g \rightarrow 0$. 

Figure~\ref{FigS3} shows the fermi level spin Berry curvature obtained from the first principles calculation. The two Berry curvature hotspots located at the valleys can be treated as independent at the vicinity of the fermi level, for which the linear approaximation is valid. This is due to the fact that the valley index is a good quantum number in this limit. The Berry curvature in each valid, then, follows Eq.~(\ref{S14}) with two peaks centered at the valleys, as explicitly shown in Fig.~\ref{FigS3}(b). 


Utilizing Eq.~(\ref{S12}), the current density becomes $J_{x}^{\eta s} = \sigma_{xy}^{\eta s}E_y$, where the conductivity is
\begin{eqnarray}
\sigma_{xy}^{\eta s}(\epsilon_F) = -\frac{e^2}{\hbar}\int \frac{d^2\textbf{q}}{(2\pi)^2}\sum_n f_{n\textbf{q}}^{\eta s}(\epsilon_F)\Omega_{n, xy}^{\eta s}(\textbf{q}).
\label{S19}
\end{eqnarray}

In the following, we will show that the spin Hall conductivity can be reversed by switching the in-plane polarization of the SnSe layer. To this end, we begin by defining the spin Hall conductivity as
\begin{eqnarray}
\sigma_{xy}^z(\epsilon_F) = \frac{\hbar}{2e}\sum_{\eta s} s \sigma_{xy}^{\eta s}(\epsilon_F).
\label{S20}
\end{eqnarray}
We focus on the spin response due to states near the Fermi level $\epsilon_F$, for which we have $s = -\eta \operatorname{sign}(\Delta_c)$. The spin Hall conductivity becomes
\begin{eqnarray}
\sigma_{xy}^{z}(\epsilon_F) = \frac{e}{2}\operatorname{sign}(\Delta_c)\int \frac{d^2\textbf{q}}{(2\pi)^2}\sum_{n\eta s} \eta f_{n\textbf{q}}^{\eta s}(\epsilon_F)\Omega_{n, xy}^{\eta s}(\textbf{q}).
\label{S21}
\end{eqnarray}

Because of time-reversal symmetry, $f_{n\textbf{q}}^{\eta s}(\epsilon_F) = f_{n\textbf{q}}^{(-\eta) (-s)}(\epsilon_F)$, where $f_{n\textbf{q}}^{\eta s}(\epsilon_F) = [1 + \exp{(\epsilon_{n}^{\eta s}(\textbf{q}) - \epsilon_F)/k_BT]}^{-1}$ is the Fermi-Dirac occupation function. Hence, the summation becomes
\begin{eqnarray}
\sum_{n\eta s} \eta f_{n\textbf{q}}^{\eta s}(\epsilon_F)\Omega_{n, xy}^{\eta s}(\textbf{q}) = \sum_{n}  f_{n\textbf{q}}^{+\downarrow}(\epsilon_F)\Omega_{n, xy}^{+\downarrow}(\textbf{q}) - f_{n\textbf{q}}^{-\uparrow}(\epsilon_F)\Omega_{n, xy}^{-\uparrow}(\textbf{q}) = 2\sum_{n}  f_{n\textbf{q}}^{+\downarrow}(\epsilon_F)\Omega_{n, xy}^{+\downarrow}(\textbf{q}),
\label{S22}
\end{eqnarray}
by taking advantage of Eq.~(\ref{S15}) and $\Omega_{n, xy}^{\eta s}(\textbf{q}) = \Omega_{n, xy}^{\eta s}(|\textbf{q}|)$ at small $\textbf{q}$, i.e., near the Fermi level. Therefore, 
\begin{eqnarray}
\sigma_{xy}^z(\epsilon_F) = \operatorname{sign}(\Delta_c)e\int \frac{d^2\textbf{q}}{(2\pi)^2}\sum_n f_{n\textbf{q}}^{+ \downarrow}(\epsilon_F)\Omega_{n, xy}^{+ \downarrow}(\textbf{q}),
\label{S23}
\end{eqnarray}
which depends on the polarization states of the SnSe layer through $\operatorname{sign}(\Delta_c)$. Thus, the spin Hall currents can be reversed by switching the ferroelectric order parameter of SnSe. The result derived here is not confined to toy model and, thus, also applied to a more rigorous first principles calculations as long as the physics at the vicinity of the Fermi level is considered. 

From the explicit valley/spin-resolved Berry curvature derived for the toy model, it is possible to obtain analytical expressions for the anomalous current density for electrons at valley $\eta$ and spin $s$. For simplicity, we examine the zero temperature conductivities in the following, i.e., the Fermi-Dirac distribution is taken to be $f_{n\textbf{q}}^{\eta s}(\epsilon_F) = \Theta (\epsilon_F - \epsilon_{ns}^{\eta}(\textbf{q}))$, where $\Theta$ is the Heaviside function. The integration can be performed very easily and renders the important results
\begin{eqnarray}
\sigma_{xy}^{+ [\operatorname{sign}(\Delta)]}(\epsilon_F) = -\sigma_{xy}^{- [-\operatorname{sign}(\Delta)]}(\epsilon_F) = \Tilde{\sigma}_{xy}(\epsilon_F),
\label{S20}
\end{eqnarray}
where
\begin{eqnarray}
\Tilde{\sigma}_{xy}(\epsilon_F) = \left\{ 
\begin{tabular}{cc}
    $\displaystyle  -\frac{e^2}{h}\frac{\Delta_c}{|\epsilon_F|}$, & if $|\epsilon_F| > E_g/2$ \\
     $\displaystyle  -\frac{e^2}{h}\frac{1}{2} $, & if $|\epsilon_F| < E_g/2$ 
\end{tabular}
\right., 
\label{S21}
\end{eqnarray}
for the gapped bands.


\section*{Screening length}\label{note4}
In addressing the feasibility of switching the in-plane polarization with external electric fields, it is necessary to account for the screening effects due to itinerant charges. Because the $\alpha$-Bi/SnSe is a semimetal, however, we expect large screening lengths, as we address here. The Thomas-Fermi dielectric function, $\varepsilon$, is  
\begin{eqnarray}
\varepsilon(|\textbf{q}|) = \displaystyle \kappa + \frac{e^2}{2\epsilon_0 |\textbf{q}|} D(\epsilon_F), 
\label{S22}
\end{eqnarray}
for the semimetallic bands obatined from Eq.~(\ref{S11}) with $\Delta = \Delta_c$ and $\operatorname{sign}(\Delta_c) + \eta s = 0$. The density of states at the fermi level $\epsilon_F$ can be expressed in terms of the carrier concentration $n$ as
\begin{eqnarray}
D(\epsilon_F) = \displaystyle \frac{1}{\hbar}\sqrt{\frac{n}{2\pi v_x v_y}}, 
\label{S23}
\end{eqnarray}
where $\kappa$ is the dieletric constant, from where the Thomas-Fermi screening length is obtained as
\begin{eqnarray}
\lambda_{\textrm{TF}} = \displaystyle \frac{2\epsilon_0 \kappa \hbar}{e^2} \sqrt{\frac{2\pi v_xv_y}{n}}. 
\label{S24}
\end{eqnarray}

Assuming a dielectric constant of $\kappa = 3$ and typical 2D semimetal carrier concentrations of $n = 10^{11}$ cm$^{-2}$, we obtain $\lambda_{\textrm{TF}} \approx 6$~nm. 


\section*{Charge-Spin Interconversion}\label{note4}
The charge-spin interconversion coefficients were computed from linear response theory. We utilize the Kubo formula fashioned after Smrcka-Streda~\cite{kubo1, kubo2, kubo3, kubo4}:
\begin{eqnarray}
     & \delta O_{\alpha \beta}^{\gamma} = \displaystyle \int \frac{d\textbf{k}}{(2\pi)^2} [\delta O_{\alpha \beta}^{\gamma, I}(\textbf{k}) + \delta O_{\alpha \beta}^{\gamma, II}(\textbf{k})],
     \label{eq2}
\end{eqnarray}
with integrands 
\begin{subequations}
\begin{equation}
\delta O_{\alpha \beta}^{{\gamma}, I}(\textrm{k}) = \displaystyle -\frac{e \hbar}{\pi} {\Gamma}^2 \sum_{nm} \frac{\operatorname{Re}
[\langle n\textbf{k}|\hat{O}_{\alpha}^{\gamma}|m\textbf{k} \rangle \langle m\textbf{k}| \hat{v}_{\beta}| n \textbf{k}\rangle]}{[({\epsilon}_{F} - {\epsilon}_{n{\textbf{k}}})^2+{\Gamma}^2][({\epsilon}_{F} - {\epsilon}_{m{\textbf{k}}})^2+{\Gamma}^2]},
\label{eq3a}
\end{equation}
\begin{equation}
\delta O_{\alpha \beta}^{\gamma, II}(\textrm{k}) = \displaystyle -2e\hbar \sum_{n,m\neq n}(f_{n\textbf{k}} - f_{m\textbf{k}}) \frac{\operatorname{Im}[\langle n\textbf{k}|\hat{O}_{\alpha}^{\gamma}|m\textbf{k} \rangle \langle m\textbf{k}| \hat{v}_{\beta}| n\textbf{k}\rangle]}{({\epsilon}_{n\textbf{k}} - {\epsilon}_{m\textbf{k}})^2  + {\Gamma}^2},
\label{eq3b}
\end{equation}
\end{subequations}
valid in the weak disorder limit described by a constant band broadening $\Gamma$, where $\hat{v}_{\beta}$ is the $\beta = x,y$ component of the velocity operator, $\hat{O}_{\alpha}^{\gamma}$ is the perturbed physical observable with spin index $\gamma = x,y,z$ and $| n\textbf{k}\rangle$ is the eigenstate associated with the band $\epsilon_{n\textbf{k}}$ of the unperturbed system. 
The charge and spin Hall responses are obtained through $\hat{O}_{\alpha}^{\gamma} \rightarrow -e\hat{v}_{\alpha}$, $\hat{O}_{\alpha}^{\gamma} \rightarrow (2/\hbar)\hat{Q}_{\alpha}^{\gamma}$, where the spin current operator is defined as $\hat{Q}_{\alpha}^{\gamma} = (1/2)\{\hat{s}^{\gamma}, \hat{v}_{\alpha}\}$.

\section*{First principles calculations}\label{note5}

\begin{figure}[t]
\centerline{\includegraphics[width = \linewidth]{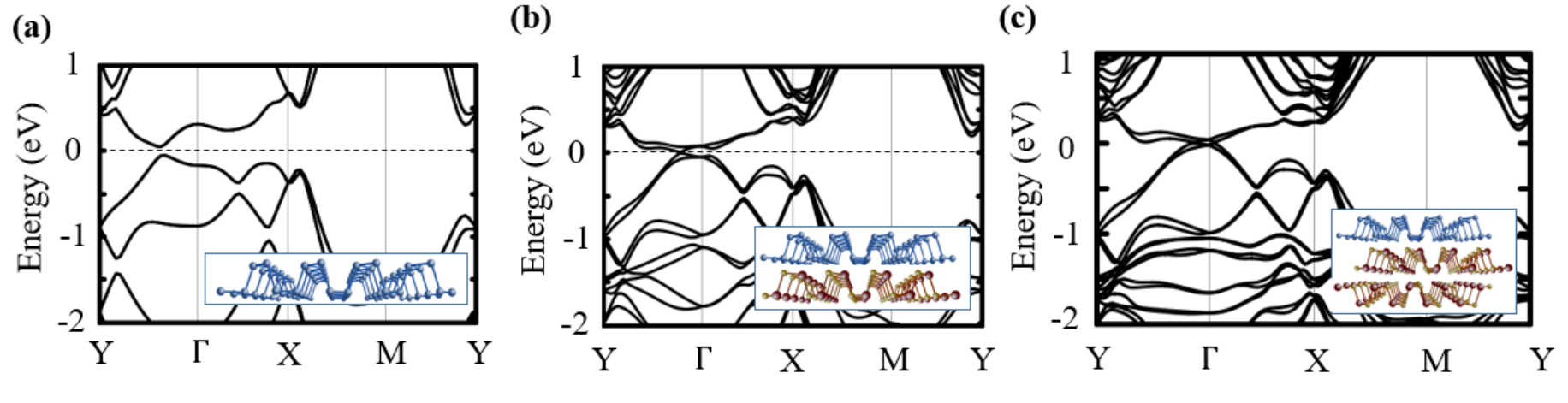}}
\caption{Band structure of (a) monolayer $\alpha$-Bi, (b) $\alpha$-Bi/SnSe van der Waals heterostructure and (c) Bi~(100)/2L-SnSe. Horizontal dashed line demarcates the fermi level. }
\label{FigS5}
\end{figure}

The first-principles calculations based on density functional theory (DFT)~\cite{Kohn1965} was carried by out Vienna \textit{ab initio} simulation package (VASP)~\cite{Kresse1996}. 
The exchange-correlation (XC) functional was treated within the generalized gradient approximation of Perdew-Burke-Ernzerhof (PBE)~\cite{Perdew1996} with noncollinear spin polarization.~\cite{PhysRevB.93.224425}
The electronic wavefunctions were expanded by planewave basis with kinetic energy cutoff of 500~eV. We employed the projector-augmented wave pseudopotentials{~\cite{{Blochl1994},{Kresse1999}}} to describe the valence electrons, and Grimme-D3 van der Waals correction~\cite{grimme2010consistent} was chosen to describe interlayer interaction.

The experimental values of lattice constants were used for $\alpha$-Bi,($a_\textrm{armchair}=4.84$~{\AA}, $b_\textrm{zigzag}=4.49~${\AA})~\cite{lu2023observation} and tensile strain was applied to SnSe substrate to make commensurate $\alpha$-Bi/SnSe heterostructure. 
The sufficiently large vacuum region ($>$20~{\AA}) was included in the periodic cell to mimic two-dimensional layered structure. The crystal structure of monolayer $\alpha$-Bi was fully realxed while maintaining $Cmce$ symmetry. In the $\alpha$-Bi/SnSe heterostructure, only SnSe was fully realxed to avoid buckled phase of $\alpha$-Bi layer.~\cite{Lu2015}

The spontaneous polarization $\textbf{P}$ of $\alpha$-Bi/SnSe heterostructure along the armchair direction was calculated in the framework of the modern theory of polarization through Berry phase.~\cite{spaldin2012beginner}
In the modern theory of polarization, $\textbf{P}$ can be rigorously defined by adiabatic Wannier charge center change along an arbitrary path. Therefore, the absolute value of $\textbf{P}$ of $\alpha$-Bi/SnSe heterostructure is only numerically available when we have its centerosymmetric (or in-plane mirror symmetric) counterpart which guarantees $\textbf{P}=0$. However, due to the geometry of the heterostructure, it is unfeasible to find centerosymmetric counterpart through the experimentally reasonable reaction path from $-\textbf{P}$ to $\textbf{P}$. To detour this problem, we evaluate $\textbf{P}$ of $\alpha$-Bi/SnSe heterostructure through following steps:
\begin{itemize}
\item Constructing structure of $-\textbf{P}$ using mirror operation. (Fig. 1(a) in the main manuscript) Strictly speaking, it corresponds to $\textbf{P}$ to $-\textbf{P}$ transition of SnSe and in-plane sliding of $\alpha$-Bi. 
\item Initializing a total of 11 structures along the path from $-\textbf{P}$ to $\textbf{P}$ using linear interpolation.
\item Finding minimum energy path using the nudged elastic band (NEB) calculation. In the NEB calculation, only SnSe was fully realxed due to the same reason described above.
\item Evaluating $\textbf{P}$ of each structure. Note that, $\textbf{P}$ of some structures in the reaction path were unavailable because they were calculated to be metallic due to $\alpha$-Bi. In this case, we determined $\textbf{P}$ of each structure through linear interpolation. 
\item Then, we set $\textbf{P}$ of 6th structure (in-plane mirror, Fig. 1(a) in the main manuscript) to be zero, and we obtained $\textbf{P}$ of $\alpha$-Bi/SnSe heterostructure.
\end{itemize}

\bibliography{my.bib}